\documentclass[a4paper,twocolumn,11pt]{quantumarticle}

\pdfoutput=1
\usepackage[utf8]{inputenc}
\usepackage[english]{babel}
\usepackage[T1]{fontenc}
\usepackage{amsmath,amssymb,amsfonts,bm}
\usepackage{graphicx}

\graphicspath{{./}{figures/}}
\usepackage{braket}
\usepackage{xcolor}
\usepackage{tikz}
\usetikzlibrary{quantikz}
\usepackage{booktabs}
\usepackage{url}
\usepackage{hyperref}
\usepackage{mathtools}
\usepackage{enumitem}

\hypersetup{
  colorlinks=true,
  linkcolor=blue,
  citecolor=blue,
  urlcolor=blue
}

\newcommand{\tr}{\operatorname{Tr}}

\newcommand{\dd}{\mathrm{d}}

\newcommand{\Mmsg}{M_{\rm msg}}
\newcommand{\Mdump}{M_{\rm dump}}
\newcommand{\backend}[1]{\texttt{\detokenize{#1}}}

\begin{document}

\title{Closed Timelike Curve Decoding on Quantum Hardware}

\author{Sai Nandan Morapakula}
\email{s.morapakula001@umb.edu}
\affiliation{Department of Physics, University of Massachusetts Boston, Boston, MA 02125, USA}
\orcid{0009-0002-5391-3201}
\author{Kazuki Ikeda}
\email{kazuki.ikeda@umb.edu}
\orcid{0000-0003-3821-2669}
\affiliation{Department of Physics, University of Massachusetts Boston, Boston, MA 02125, USA}
\affiliation{Center for Nuclear Theory, Department of Physics and Astronomy, Stony Brook University, Stony Brook, New York 11794-3800, USA}
\maketitle

\begin{abstract}
Deutsch closed timelike curves (D-CTCs) are described by a fixed-point condition for a chronology-violating register.  We study a finite-dimensional circuit model that places a Hayden--Preskill/Yoshida--Kitaev recovery map inside such a consistency loop.  A register-routing construction makes the Deutsch map explicit: an initial SWAP moves the incoming CTC state to an idle dump register, the scrambler and decoder act on the remaining active registers, and a final SWAP writes the recovered message back to the CTC register.  When the active branch recovers the message, the induced map on the CTC register is the replacement channel \(\sigma\mapsto \rho_M\), with the unique fixed point \(\rho_M\).  We implement the associated Lloyd-type post-selected decoder circuits on quantum hardware and formulate a classical-feedback iteration for the experimentally estimated map.  Qiskit simulations and IBM-hardware data for single-qubit instances quantify decoder fidelity, post-selection overhead, routing-dependent noise, and quantum-geometric susceptibility.
\end{abstract}

\section{Introduction}
\begin{figure}[t]
\centering
\includegraphics[width=\linewidth]{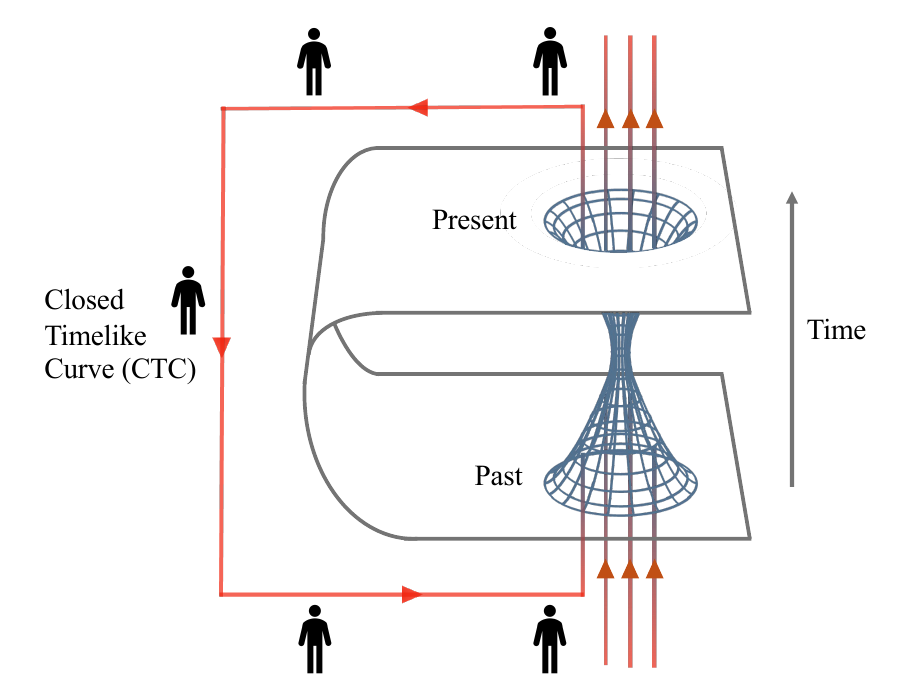}
\caption{Schematic illustration of a closed timelike curve (CTC).  A worldline traversing a wormhole-like geometry connects the present to its own past, forming a closed loop in spacetime.  The figure motivates the finite-dimensional quantum-information models studied below.}
\label{fig:ctc-schematic}
\end{figure}

Closed timelike curves (CTCs) are hypothetical spacetime structures that occur in some solutions of general relativity \cite{Godel1949,Gott1991,Morris1988}; Fig.~\ref{fig:ctc-schematic} gives a schematic illustration.  If a physical system could interact with its own past, the theory would have to address consistency puzzles such as the grandfather paradox \cite{Lewis2016}.  Deutsch proposed a quantum model in which the state of a chronology-violating register is required to be a fixed point of the completely positive trace-preserving (CPTP) map induced by its interaction with chronology-respecting registers \cite{Deutsch1991,Pienaar2013}.  This rule avoids logical contradictions at the level of the density matrix, but it also makes the effective input--output map on chronology-respecting systems nonlinear.  In the circuit model this nonlinearity has strong computational consequences: polynomial-size Deutsch-CTC circuits can decide all problems in PSPACE \cite{AaronsonWatrous2009, AaronsonEtAl2024, Brun2003, Bennett2009}, and related nonlinear or postselected models can distinguish non-orthogonal states \cite{BrunWilde2011,BrunWildeWinter2013}.

A separate line of work concerns black-hole information recovery.  In the Hayden-Preskill setting, an old black hole is modeled as a scrambler, and a small message can be recovered from the early radiation and a modest amount of late radiation \cite{HaydenPreskill2007}.  Yoshida and Kitaev gave explicit recovery procedures involving the inverse, transpose, or complex-conjugate scrambler, with deterministic versions obtained by Grover-type amplitude amplification \cite{YoshidaKitaev2017}.

Here we combine these two ideas at circuit level.  A Yoshida-Kitaev-type decoder is embedded in a Deutsch consistency loop, and the resulting fixed-point structure is controlled by the register routing.  In the ideal circuit of Fig.~\ref{fig:register-loop}, the first SWAP isolates the incoming CTC state on a spectator wire, the active decoder sees an input independent of that state, and the final SWAP writes the recovered message back into the CTC register.  Thus message recovery on the active branch induces the replacement channel \(\sigma\mapsto\rho_M\) on the CTC register and gives the unique Deutsch fixed point \(\sigma_*=\rho_M\).

This structure leads to two complementary parts of the paper.  Section~\ref{sec:deutsch-loop} proves the ideal Deutsch fixed-point statement.  The hardware sections implement a Lloyd-type post-selected decoder branch \cite{Lloyd2011PRL,Ringbauer2014,Lloyd2011PRD,Huang2026}, while Sec.~\ref{sec:hardware} also formulates a classical-feedback iteration of an estimated CPTP map.  Together these constructions connect the fixed-point theory to experimentally accessible decoder observables.  Related uses of postselection appear in the final-state projection proposal for black-hole unitarity~\cite{HorowitzMaldacena2004} and in analyses of postselected boundary conditions~\cite{GottesmanPreskill2004}.

As diagnostics, we use not only message fidelity and post-selection probability but also perturbative and quantum-geometric measures. Scrambling is commonly probed through perturbation sensitivity, out-of-time-order correlators, and related chaos diagnostics~\cite{Maldacena2016, Leone2021, Mi2021, YoshidaYao2019, Zhuang2019}. The quantum geometric tensor (QGT) provides a complementary way to quantify how rapidly a parameterized family of output states changes in projective Hilbert space~\cite{Provost1980, Kolodrubetz2017, Cheng2013, Kang2024, Austrich2022}.

Several closely related works provide context for this construction.
Bishop \textit{et al.} developed a
weak-measurement tomography framework for operationally assigning
states to chronology-violating systems in D-CTC and P-CTC models
\cite{Bishop2025}. Huang \textit{et al.} interpreted the
probabilistic Yoshida--Kitaev decoder as a postselected CTC and
demonstrated scrambled-information recovery on Quantinuum and IBM
processors \cite{Huang2026}. Building on these results, we place the decoder inside
an explicit Deutsch consistency loop and determine the induced CPTP map
on the chronology-violating register. The register-separated circuit
routes the incoming CTC state to an idle dump register; whenever the
active branch recovers the message, the Deutsch map is the replacement
channel $\sigma\mapsto\rho_M$ and has the unique fixed point
$\sigma^\ast=\rho_M$.

The rest of the paper is organized as follows.  Section~\ref{sec:deutsch-loop} states the register-separated replacement-channel theorem.  Section~\ref{sec:deterministic} reviews the deterministic decoder as a unitary dilation and clarifies the single-qubit resource scaling.  Section~\ref{sec:hardware} presents the post-selected hardware emulation and the classical-feedback formulation.  Section~\ref{sec:numerics} reports parameter sweeps, local perturbation tests, and quantum-geometric diagnostics.  The appendices give process-diagnostic data, decoder details, post-selection data, and a hardware-accessible Loschmidt-echo proxy for diagonal QGT components.

\section{Looped decoder in a Deutsch CTC}
\label{sec:deutsch-loop}

\subsection{Registers and ideal circuit}

We first consider a single-qubit message and later comment on the \(m\)-qubit generalization.  The ideal register-separated construction uses the following registers:
\begin{itemize}[leftmargin=*]
\item \(\Mmsg\): a chronology-respecting message register prepared in an arbitrary state \(\rho_M\).  After the first SWAP, the same physical wire is denoted \(\Mdump\), because it carries the incoming CTC state \(\sigma\) in the ideal proof.
\item \(C\): the CTC register.  Its incoming leg carries an a priori unknown state \(\sigma\).
\item \(E,R\): early and recent radiation subsystems.
\item \(A,A'\): ancillas prepared in a Bell state \(|\Phi^+\rangle_{AA'}\).
\item \(G\): decoder workspace.
\item \(Y\): the output register, identified with \(A'\).
\end{itemize}

The Deutsch fixed-point theorem uses a register-separated ledger.  After the first SWAP, \(C\) carries \(\rho_M\), while \(\Mdump\) carries the incoming CTC state \(\sigma\).  The active decoder acts trivially on \(\Mdump\), and all Bell-pair halves and workspaces are separate clean ancillas.  In the compressed seven-qubit circuits introduced later, the would-be CTC input is initialized; the corresponding wire can then be reused as an auxiliary register for hardware efficiency.

\begin{figure*}[t]
    \centering
\includegraphics[width=0.85\linewidth]{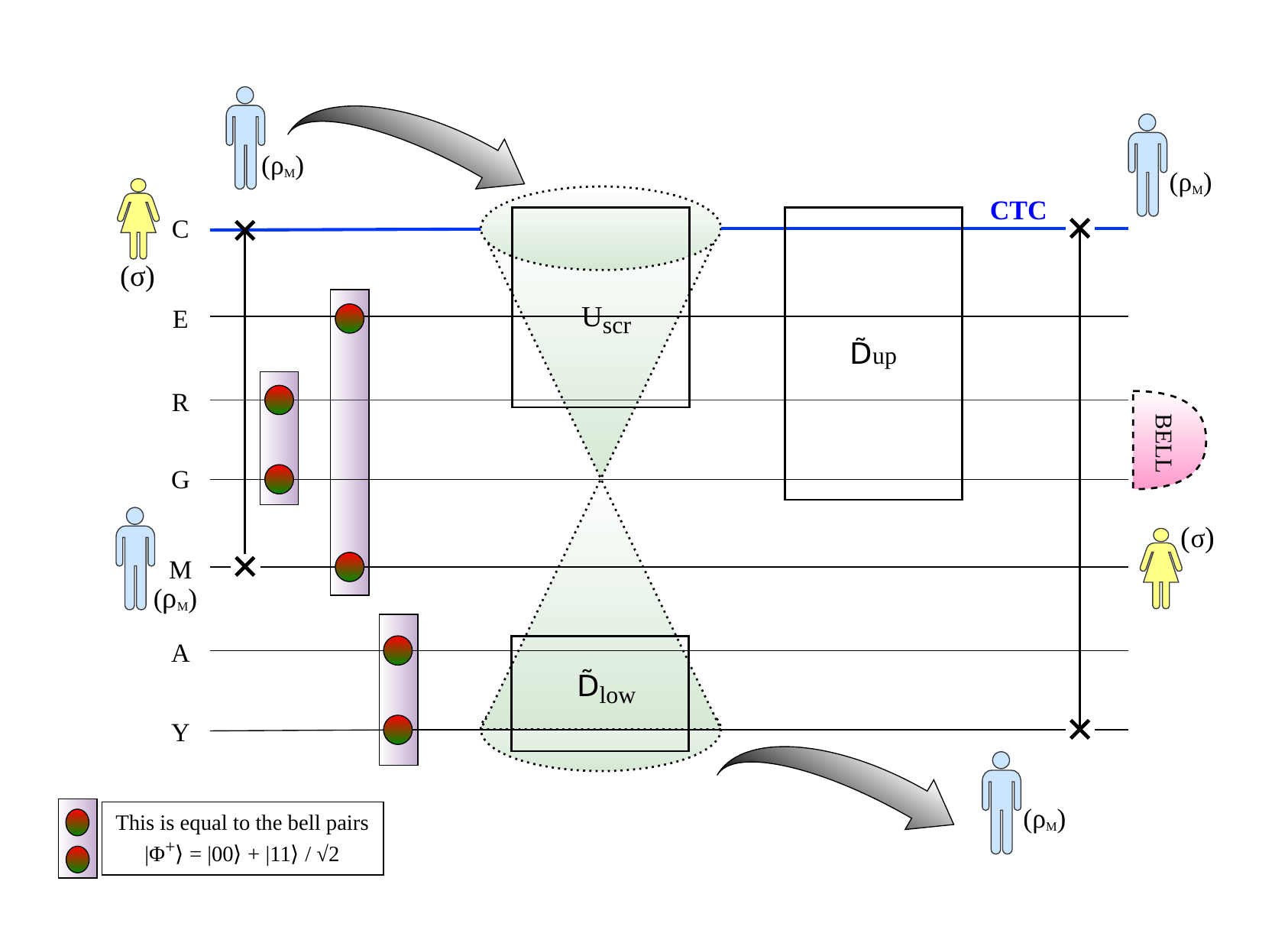}
    \caption{Register-separated ideal Deutsch-loop construction in the register order
\(C,E,R,G,M,A,Y\).
The first SWAP transfers the prepared message \(\rho_M\) from \(M\) to \(C\)
and routes the incoming CTC state \(\sigma\) to the dump wire
\(M_{\rm dump}\), which remains a spectator thereafter.
The scrambler \(U_{\rm scr}\) acts on \(C,E,R\), spreading the message
information across these registers.
The aligned blocks \(\widetilde{D}_{\rm up}\) and
\(\widetilde{D}_{\rm low}\) are the upper and lower graphical components
of the same decoder dilation and together recover the message on \(Y\).
Algebraically, the active operation has the factorized form
\(I_{M_{\rm dump}}\otimes W_{C,E,R,G,A,Y}\).
If the active decoder recovers $\rho_M$ on $Y$, the final SWAP
transfers this state to $C$, producing the replacement-channel action
$\sigma\mapsto\rho_M$.}
    \label{fig:register-loop}
\end{figure*}

Fig.~\ref{fig:register-loop} shows the ideal looped construction in the register order \(C,E,R,G,M,A,Y\).  The scrambler \(U_{\rm scr}\) acts on \(C,E,R\).  The deterministic decoder dilation is a joint active operation on \(C,E,R,G,A,Y\),
\begin{equation}
  W_{\rm act}\equiv W_{C,E,R,G,A,Y} .
  \label{eq:wact}
\end{equation}
In the diagram, this operation is drawn as two aligned graphical pieces ($\widetilde D_{\rm up}$, $\widetilde D_{\rm low}$) because the idle wire \(\Mdump\) sits between the upper and lower active wires in the chosen ordering.  Algebraically, the decoder dilation factorizes as
\begin{equation}
  W = I_{\Mdump}\otimes W_{\rm act} .
  \label{eq:factorized-W}
\end{equation}
Since the scrambler also acts trivially on \(\Mdump\), the complete operation between the two SWAPs is
\begin{equation}
\begin{aligned}
  \mathcal{A}&\equiv W U_{\rm scr}
  =I_{\Mdump}\otimes\mathcal{A}_{\rm act},\\
  \mathcal{A}_{\rm act}
  &:=W_{\rm act}\bigl(U_{\rm scr}^{CER}\otimes I_{GAY}\bigr).
\end{aligned}
  \label{eq:factorized-active}
\end{equation}
The final SWAP writes the state recovered on \(Y\) into \(C\).  Thus successful recovery by \(\mathcal{A}_{\rm act}\) becomes the replacement action \(\sigma\mapsto\rho_M\) on the CTC register.

\subsection{Overall unitary and Deutsch map}

Let \(V\) denote the global unitary on all registers,
\begin{equation}
\begin{aligned}
  V&:=\mathrm{SWAP}_{Y\leftrightarrow C}\,
  \mathcal{A}\,\mathrm{SWAP}_{\Mmsg\leftrightarrow C}\\
  &=\mathrm{SWAP}_{Y\leftrightarrow C}\,
  W U_{\rm scr}\,\mathrm{SWAP}_{\Mmsg\leftrightarrow C}.
\end{aligned}
  \label{eq:global-V}
\end{equation}
The factorization in Eq.~\eqref{eq:factorized-active} isolates the incoming CTC state from the scrambler--decoder block and turns recovery on \(Y\) into a replacement channel on \(C\).

Let \(\mathrm{CR}\) denote the tensor product of all registers except \(C\).  For a fixed chronology-respecting preparation \(\rho_{\mathrm{CR}}\), the circuit induces a linear CPTP map on the CTC register,
\begin{equation}
  \Phi_{\rho_{\mathrm{CR}}}(\sigma)
  := \tr_{\mathrm{CR}}\!\left[V(\rho_{\mathrm{CR}}\otimes \sigma)V^\dagger\right] .
  \label{eq:deutsch-map}
\end{equation}
Deutsch's self-consistency condition is
\begin{equation}
  \sigma_* = \Phi_{\rho_{\mathrm{CR}}}(\sigma_*) .
  \label{eq:deutsch-condition}
\end{equation}
If the fixed point is not unique, Deutsch also proposed a maximum-entropy selection rule \cite{Deutsch1991}.  In the ideal construction below, the fixed point is unique, so no selection rule is needed.  Given a fixed point \(\sigma_*\), the output on chronology-respecting registers is
\begin{equation}
  \rho'_{\mathrm{CR}} = \tr_C\!\left[V(\rho_{\mathrm{CR}}\otimes \sigma_*)V^\dagger\right],
  \label{eq:cr-output}
\end{equation}
which can depend nonlinearly on \(\rho_{\mathrm{CR}}\) through the dependence of \(\sigma_*\) on \(\rho_{\mathrm{CR}}\) \cite{AaronsonWatrous2009}.


\subsection{Register-separated fixed-point theorem}

\textit{Theorem II.1 (Register-separated replacement loop).}
Consider the ideal Deutsch-loop circuit shown in Fig.~\ref{fig:register-loop}, where qubit $C$ serves as the CTC system corresponding to the schematic in Fig.~\ref{fig:ctc-schematic}. The chronology-respecting preparation is \(\rho_M\) on \(\Mmsg\) and a fixed ancilla state \(\tau\) on \(E,R,G,A,Y\), independent of the incoming CTC state \(\sigma\) on \(C\).  Suppose that the complete active operation between the two SWAPs factorizes as
\begin{equation}
\begin{aligned}
  \mathcal{A}&=W U_{\rm scr}
  =I_{\Mdump}\otimes\mathcal{A}_{\rm act},\\
  \mathcal{A}_{\rm act}
  &=W_{\rm act}\bigl(U_{\rm scr}^{CER}\otimes I_{GAY}\bigr),
\end{aligned}
  \label{eq:theorem-factor}
\end{equation}
and that this active scrambler--decoder block reproduces the message state on \(Y\),
\begin{equation}
\begin{aligned}
  &\tr_{C,E,R,G,A}\!\left[
  \mathcal{A}_{\rm act}
  \bigl((\rho_M)_C\otimes\tau_{E,R,G,A,Y}\bigr)
  \mathcal{A}_{\rm act}^\dagger
  \right]\\
  &\hspace{5em}=\rho_M .
\end{aligned}
  \label{eq:theorem-recovery}
\end{equation}
Then the induced Deutsch map on the CTC register is exactly the replacement channel
\begin{equation}
  \Phi_{\rho_M}(\sigma)=\rho_M \qquad \text{for all states } \sigma .
  \label{eq:replacement-channel}
\end{equation}
Consequently, Deutsch's consistency equation has the unique fixed point
\begin{equation}
  \sigma_* = \rho_M .
  \label{eq:unique-fixed-point}
\end{equation}

\textit{Proof.}
Let \(S_{MC}\) be the first SWAP between \(C\) and the message wire, and let \(S_{CY}\) be the final SWAP between \(C\) and \(Y\).  On the two registers affected by the first SWAP,
\begin{equation}
  S_{MC}\bigl(\sigma_C\otimes (\rho_M)_{\Mmsg}\bigr)S_{MC}^\dagger
  = (\rho_M)_C\otimes \sigma_{\Mdump} .
\end{equation}
Thus the incoming CTC state appears only on \(\Mdump\), while the active registers contain \(\rho_M\) and the fixed ancillas.  Using Eq.~\eqref{eq:theorem-factor}, the state after the full scrambler--decoder block is
\begin{equation}
  \sigma_{\Mdump}\otimes
  \mathcal{A}_{\rm act}\bigl((\rho_M)_C\otimes\tau_{E,R,G,A,Y}\bigr)
  \mathcal{A}_{\rm act}^\dagger .
\end{equation}
The reduced state on \(Y\) is independent of \(\sigma\) and equals \(\rho_M\) by Eq.~\eqref{eq:theorem-recovery}.  The final SWAP transfers this state to the outgoing CTC register.  Hence
\begin{align}
X &\equiv S_{CY}\mathcal{A}S_{MC}
   =S_{CY}W U_{\rm scr}S_{MC}, \\
\Phi_{\rho_M}(\sigma)
&=\tr_{\mathrm{CR}}\!\left[
  X\bigl(\sigma_C\otimes(\rho_M)_{\Mmsg}\otimes\tau\bigr)X^\dagger
\right] \\
&=\rho_M .
\end{align}
for every input state \(\sigma\).  The Deutsch equation \(\sigma=\Phi_{\rho_M}(\sigma)\) therefore reduces to \(\sigma=\rho_M\), whose solution is unique.\hfill\(\square\)

\subsection{Approximate replacement and noisy devices}

The fixed-point theorem is best viewed as a statement about the CTC-input map at fixed message.  In the noiseless register-separated model the map is exactly
\begin{equation}
  \Phi(\sigma)=\rho_M \qquad \forall\,\sigma .
\end{equation}
Gate noise, crosstalk, leakage, or imperfect decoding can turn this exact replacement action into an approximate one.  If the actual CTC-input map at fixed \(\rho_M\) satisfies
\begin{equation}
  \|\Phi-\mathcal{R}_{\rho_M}\|_\diamond\leq \delta,
  \label{eq:diamond-condition}
\end{equation}
where \(\mathcal{R}_{\rho_M}(\tau)=\rho_M\) for every input \(\tau\), then any fixed point of \(\Phi\) is close to \(\rho_M\).

\textit{Lemma II.2 (Approximate fixed points under approximate replacement).}
If Eq.~\eqref{eq:diamond-condition} holds and \(\sigma_*\) is a fixed point of \(\Phi\), then
\begin{equation}
  \|\sigma_* - \rho_M\|_1 \leq \delta .
  \label{eq:fixed-point-diamond-bound}
\end{equation}

\textit{Proof.}
For any state \(\tau\), \(\|\Phi(\tau)-\rho_M\|_1\leq \|\Phi-\mathcal{R}_{\rho_M}\|_\diamond\leq\delta\).  Applying this to \(\tau=\sigma_*\) and using \(\Phi(\sigma_*)=\sigma_*\) gives Eq.~\eqref{eq:fixed-point-diamond-bound}.\hfill\(\square\)

Appendix~\ref{app:qpt} reports a closely related hardware diagnostic: a message-to-output process reconstructed by varying the prepared message and measuring the recovered register.  Lemma~II.2 instead concerns the CTC-input map \(\sigma\mapsto\Phi_{\rho_{\mathrm{CR}}}(\sigma)\) at fixed message.  A direct test of the lemma would vary the incoming CTC state \(\sigma\), hold \(\rho_M\) fixed, reconstruct the induced map on \(C\), and bound its diamond-norm distance to \(\mathcal{R}_{\rho_M}\).

\section{Deterministic decoder as a unitary dilation}
\label{sec:deterministic}

In the Hayden--Preskill setting, the Yoshida--Kitaev decoder is naturally probabilistic.  The basic decoding branch succeeds only when a specified maximally entangled outcome occurs.  For an \(m\)-qubit message with dimension \(d_M=2^m\), the success probability scales as \(p_{\rm succ}\sim 1/d_M^2\) \cite{YoshidaKitaev2017}.  Grover-type amplitude amplification can make the recovery approximately deterministic using \(O(1/\sqrt{p_{\rm succ}})=O(d_M)=O(2^m)\) Grover iterates \cite{BrunWilde2011,Grover1996,Brassard2002}.

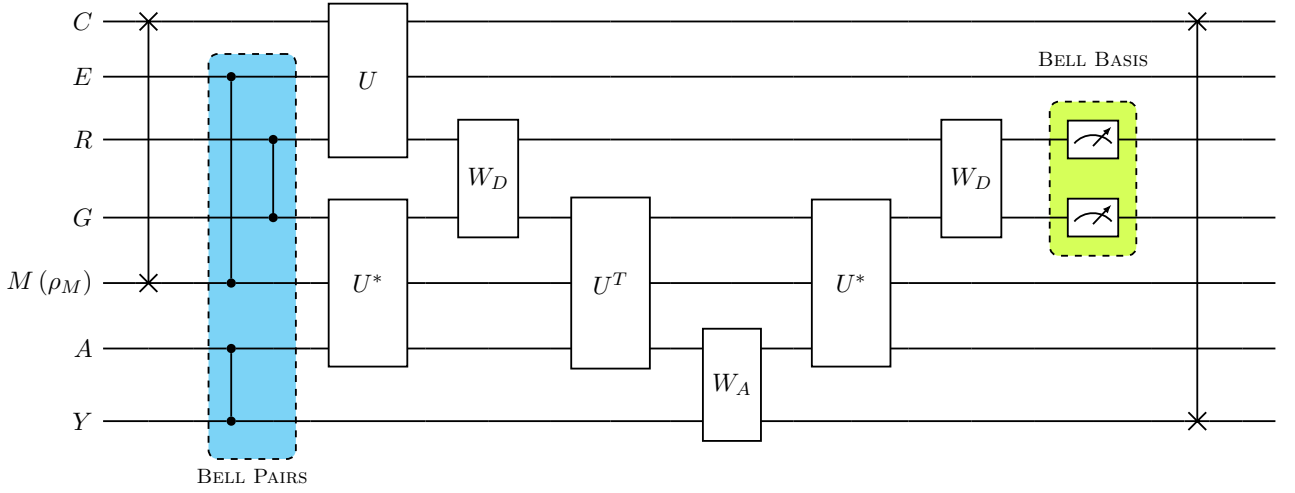
\begin{figure*}[t] 
    \centering
        \resizebox{\textwidth}{!}{%
\begin{quantikz}[color=black]
    \lstick{$C$} & \swap{4} & \qw & \qw & \qw & \qw & \gate[wires=3,style={minimum width=1.2cm}]{U} & \qw & \qw & \qw & \qw & \qw & \qw & \qw & \qw & \qw & \qw & \qw & \qw & \qw & \swap{6} & \qw & \qw \\
    \lstick{$E$} & \qw & \qw & \ctrl{3}\gategroup[6,steps=2,style={dashed,rounded corners,fill=cyan!45, inner xsep=4pt, inner ysep=4pt},background,label style={label position=below,anchor=north,yshift=-0.2cm}]{{\footnotesize \sc \color{black} Bell Pairs}} & \qw & \qw & & \qw & \qw & \qw & \qw & \qw & \qw & \qw & \qw & \qw & \qw & \qw & \qw & \qw & \qw & \qw & \qw \\
    \lstick{$R$} & \qw & \qw & \qw & \ctrl{1} & \qw & & \qw & \gate[wires=2,style={minimum width=0.8cm}]{W_D} & \qw & \qw & \qw & \qw & \qw & \qw & \qw & \gate[wires=2,style={minimum width=0.8cm}]{W_D} & \qw & \meter{}\gategroup[2,steps=1,style={dashed,rounded corners,fill=lime!65, inner xsep=4pt, inner ysep=4pt},background,label style={label position=above,anchor=south,yshift=0.2cm}]{{\footnotesize \sc \color{black} Bell Basis}} & \qw & \qw & \qw & \qw \\
    \lstick{$G$} & \qw & \qw & \qw & \control{} & \qw & \gate[wires=3,style={minimum width=1.2cm}]{U^*} & \qw & & \qw & \gate[wires=3,style={minimum width=1.2cm}]{U^T} & \qw & \qw & \qw & \gate[wires=3,style={minimum width=1.2cm}]{U^*} & \qw & & \qw & \meter{} & \qw & \qw & \qw & \qw \\
    \lstick{$M\,(\rho_M)$} & \targX{} & \qw & \control{} & \qw & \qw & & \qw & \qw & \qw & & \qw & \qw & \qw & & \qw & \qw & \qw & \qw & \qw & \qw & \qw & \qw \\
    \lstick{$A$} & \qw & \qw & \ctrl{1} & \qw & \qw & & \qw & \qw & \qw & & \qw & \gate[wires=2,style={minimum width=0.8cm}]{W_A} & \qw & & \qw & \qw & \qw & \qw & \qw & \qw & \qw & \qw \\
    \lstick{$Y$} & \qw & \qw & \control{} & \qw & \qw & \qw & \qw & \qw & \qw & \qw & \qw & & \qw & \qw & \qw & \qw & \qw & \qw & \qw & \targX{} & \qw & \qw
\end{quantikz}%
}

\caption{Single-qubit, seven-qubit amplitude-amplified decoder circuit used in the hardware study.  The first SWAP moves the prepared message \(\rho_M\) from \(M\) into \(C\).  In this hardware convention the would-be CTC input is initialized, so the post-SWAP wire labelled \(M\) is clean and can be reused inside the Bell-pair/conjugate-decoder block.  The register-separated arbitrary-\(\sigma\) version is shown in Fig.~\ref{fig:register-loop}.}
\label{fig:single-qubit-decoder}
\end{figure*}

We use two circuit forms.  In Sec.~\ref{sec:deutsch-loop}, the decoder is an ideal unitary block in a register-separated Deutsch loop, where \(\Mdump\) remains idle.  In the single-qubit hardware study, Fig.~\ref{fig:single-qubit-decoder} gives a compressed seven-qubit amplitude-amplified decoder circuit.  Since the would-be CTC input is initialized in the hardware run, the post-SWAP wire labelled \(M\) is clean and can be reused as a decoder auxiliary wire.  This is the circuit form used for decoder-fidelity and post-selection measurements; the arbitrary-\(\sigma\) Deutsch map is represented by the register-separated circuit of Fig.~\ref{fig:register-loop}.

For the single-qubit instance, one Grover iterate is sufficient.  More generally, the required number of iterates grows exponentially with \(m\), as summarized in Appendix~\ref{app:det-decoder}.  Resource estimates for larger messages must therefore include the amplitude-amplification depth as well as the gate count of a single iterate.

\section{Emulating the decoder loop on quantum hardware}
\label{sec:hardware}

Gate-model quantum processors implement linear CPTP maps.  Deutsch CTCs lead to an effectively nonlinear map on chronology-respecting systems because the fixed point can depend on the input.  We use three complementary descriptions:
\begin{itemize}[leftmargin=*]
\item \textbf{Ideal Deutsch loop:} a fixed-point construction for a CPTP map on \(C\).  In the circuit model this layer can have PSPACE computational power \cite{AaronsonWatrous2009}.
\item \textbf{Lloyd-type post-selected emulation:} a linear quantum circuit followed by conditioning on a specified measurement outcome \cite{Lloyd2011PRL}.  This is related to PostBQP \cite{Aaronson2005} and can be exponentially costly in the inverse success probability.
\item \textbf{Classical-feedback iteration:} an experimentally estimated CPTP map is iterated in classical control until a fixed point is reached or until a stopping criterion is met.
\end{itemize}

The seven-qubit circuits used below realize the post-selected decoder branch.  The classical-feedback construction is formulated as an iterative use of the experimentally estimated map.  Both use initialized circuit inputs, while the register-separated theorem of Sec.~\ref{sec:deutsch-loop} supplies the Deutsch fixed-point statement.

The post-selection overhead scales as
\begin{equation}
  N_{\rm total}=\frac{N_{\rm useful}}{p_{\rm succ}}
  \sim 4^m N_{\rm useful},
  \label{eq:postselection-cost}
\end{equation}
for an \(m\)-qubit message, because \(p_{\rm succ}\sim 4^{-m}\) \cite{YoshidaKitaev2017}.  For \(m=1\), this is a factor of about four.  For larger messages, strict post-selection becomes exponentially costly unless it is replaced by amplitude amplification, feedforward, or another deterministic recovery method.

\subsection{Post-selected decoder emulation}

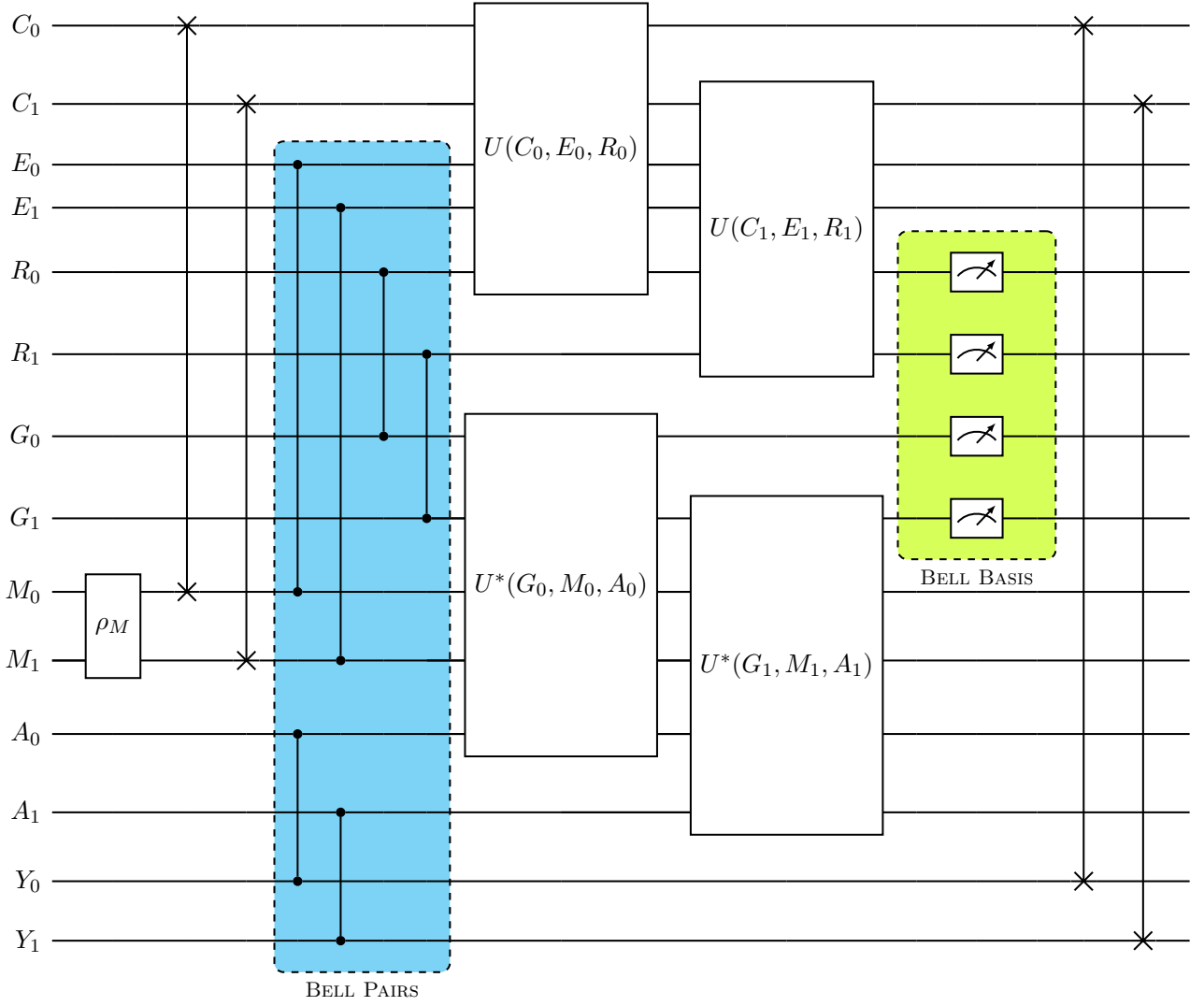
\begin{figure*}[t] 
    \centering
    \resizebox{\textwidth}{!}{%
\begin{quantikz}[color=black]
    \lstick{$C_0$} & \qw & \swap{8} & \qw & \qw & \qw & \qw & \qw & \gate[wires=5,style={minimum width=1.8cm}]{U(C_0, E_0, R_0)} & \qw & \qw & \qw & \qw & \swap{12} & \qw & \qw \\
    \lstick{$C_1$} & \qw & \qw & \swap{8} & \qw & \qw & \qw & \qw & \qw & \gate[wires=5,style={minimum width=1.8cm}]{U(C_1, E_1, R_1)} & \qw & \qw & \qw & \qw & \swap{12} & \qw \\
    \lstick{$E_0$} & \qw & \qw & \qw & \ctrl{6}\gategroup[12,steps=4,style={dashed,rounded corners,fill=cyan!45, inner xsep=4pt, inner ysep=4pt},background,label style={label position=below,anchor=north,yshift=-0.2cm}]{{\footnotesize \sc \color{black} Bell Pairs}} & \qw & \qw & \qw & \qw & \qw & \qw & \qw & \qw & \qw & \qw & \qw \\
    \lstick{$E_1$} & \qw & \qw & \qw & \qw & \ctrl{6} & \qw & \qw & \qw & \qw & \qw & \qw & \qw & \qw & \qw & \qw \\
    \lstick{$R_0$} & \qw & \qw & \qw & \qw & \qw & \ctrl{2} & \qw & \qw & \qw & \qw \gategroup[4,steps=3,style={dashed,rounded corners,fill=lime!65, inner xsep=4pt, inner ysep=4pt},background,label style={label position=below,anchor=north,yshift=-0.2cm}]{{\footnotesize \sc \color{black} Bell Basis}} & \meter{} & \qw & \qw & \qw & \qw \\
    \lstick{$R_1$} & \qw & \qw & \qw & \qw & \qw & \qw & \ctrl{2} & \qw & \qw & \qw & \meter{} & \qw & \qw & \qw & \qw \\
    \lstick{$G_0$} & \qw & \qw & \qw & \qw & \qw & \control{} & \qw & \gate[wires=5,style={minimum width=1.8cm}]{U^*(G_0, M_0, A_0)} & \qw & \qw & \meter{} & \qw & \qw & \qw & \qw \\
    \lstick{$G_1$} & \qw & \qw & \qw & \qw & \qw & \qw & \control{} & \qw & \gate[wires=5,style={minimum width=1.8cm}]{U^*(G_1, M_1, A_1)} & \qw & \meter{} & \qw & \qw & \qw & \qw \\
    \lstick{$M_0$} & \gate[wires=2]{\rho_M} & \targX{} & \qw & \control{} & \qw & \qw & \qw & \qw & \qw & \qw & \qw & \qw & \qw & \qw & \qw \\
    \lstick{$M_1$} & \qw & \qw & \targX{} & \qw & \control{} & \qw & \qw & \qw & \qw & \qw & \qw & \qw & \qw & \qw & \qw \\
    \lstick{$A_0$} & \qw & \qw & \qw & \ctrl{2} & \qw & \qw & \qw & \qw & \qw & \qw & \qw & \qw & \qw & \qw & \qw \\
    \lstick{$A_1$} & \qw & \qw & \qw & \qw & \ctrl{2} & \qw & \qw & \qw & \qw & \qw & \qw & \qw & \qw & \qw & \qw \\
    \lstick{$Y_0$} & \qw & \qw & \qw & \control{} & \qw & \qw & \qw & \qw & \qw & \qw & \qw & \qw & \targX{} & \qw & \qw \\
    \lstick{$Y_1$} & \qw & \qw & \qw & \qw & \control{} & \qw & \qw & \qw & \qw & \qw & \qw & \qw & \qw & \targX{} & \qw
    \end{quantikz}%
}
\caption{Probabilistic Yoshida--Kitaev decoder used as a Lloyd-type post-selected emulation for a multi-qubit message.  The would-be CTC input is initialized in this compressed circuit, so the post-SWAP wires labelled \(M_0,M_1\) can be used as auxiliary decoder wires.  Runs are retained when the Bell measurements on \((R_0,G_0)\) and \((R_1,G_1)\) give the selected \(|\Phi^+\rangle\) outcomes.  The register-separated Deutsch replacement-channel circuit is the ideal construction in Fig.~\ref{fig:register-loop}.}
\label{fig:probabilistic-decoder}
\end{figure*}

We implemented the post-selected decoder-emulation circuit shown in Fig.~\ref{fig:probabilistic-decoder}.  The message state
\begin{equation}
\begin{aligned}
  |\psi_M\rangle &= U(\theta_m,\phi_m,0)|0\rangle,\\
  (\theta_m,\phi_m)&=(2.5,2.0).
\end{aligned}
  \label{eq:message-state}
\end{equation}
was prepared on register \(M\), swapped into \(C\), scrambled by \(U\) on \((C,E,R)\), and decoded in the compressed emulator by the conjugate decoder block.  Because the would-be CTC input is initialized in this implementation, the post-SWAP wire labelled \(M\) can be used as an auxiliary wire.  The register-separated proof keeps this wire as \(\Mdump\), while the hardware circuit uses the initialized branch to reduce the qubit count.

The scrambler used in the experiments is the real self-inverse unitary given in Appendix~\ref{app:det-decoder}.  For this special unitary, \(U=U^*=U^T=U^\dagger\); nevertheless, we keep the inverse, transpose, and complex-conjugate notation in the circuit descriptions because the distinction matters for a general scrambler.  For each Pauli measurement basis \((X,Y,Z)\) on the output register, raw bitstrings were stored and post-processed classically.  A run is retained only when the Bell measurement gives the selected \(|\Phi^+\rangle\) outcome.

\begin{figure}[t]
\centering
\includegraphics[width=\linewidth]{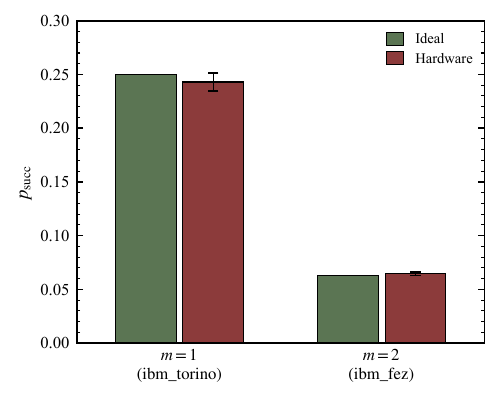}
\caption{Post-selection probability for the one- and two-qubit decoder-emulation circuits.  The ideal values are \(p_{\rm succ}=1/4\) for \(m=1\) and \(p_{\rm succ}=1/16\) for \(m=2\).  The hardware estimates are consistent with these Bell-projection probabilities within the displayed confidence intervals, confirming the expected sampling overhead for the selected branch.}
\label{fig:postselection-rate}
\end{figure}

Fig.~\ref{fig:postselection-rate} shows that, for \(m=1\), the measured post-selection probability is compatible with the ideal value \(1/4\).  For \(m=2\), the corresponding ideal value is \(1/16\).  Agreement of these probabilities verifies the Bell-projection statistics and the sampling overhead; the recovered-state fidelity is measured separately by output tomography.

\subsection{Classical-feedback fixed-point iteration}

A fixed point of the estimated map can also be obtained by classical feedback.  At each iteration, the device serves as a black-box evaluator of the linear map
\begin{equation}
  \Phi(\sigma):=\tr_{\mathrm{CR}}\!\left[V(\rho_{\mathrm{CR}}\otimes\sigma)V^\dagger\right].
  \label{eq:classical-feedback-map}
\end{equation}
A classical controller prepares \(\sigma_k\), estimates \(\sigma_{k+1}=\Phi(\sigma_k)\), and repeats.  If \(\Phi\) is a contraction in trace norm,
\begin{equation}
  \|\Phi(\rho)-\Phi(\sigma)\|_1 \leq \eta\|\rho-\sigma\|_1,
  \qquad 0\leq\eta<1,
  \label{eq:contraction}
\end{equation}
then Banach's fixed-point theorem guarantees a unique fixed point and geometric convergence.

In practice, convergence is monitored through the trace-norm step size \(s_k=\|\sigma_{k+1}-\sigma_k\|_1\).  For a single-qubit CTC register, writing \(\sigma_k=(I+\bm r_k\cdot\bm\sigma)/2\), one has \(\|\sigma_{k+1}-\sigma_k\|_1=|\bm r_{k+1}-\bm r_k|\).  The trace distance in the standard convention would be half of this value.  Three-basis Pauli tomography estimates the Bloch vector with shot cost \(O(\epsilon^{-2})\) per iteration.  For larger registers, full tomography scales exponentially.  Classical shadows \cite{Huang2020} can estimate many selected observables efficiently, while efficient trace-norm estimation requires additional structure such as low rank, locality, or a restricted observable class.

Suppose the iteration stops after step \(k_*\) because
\begin{equation}
  s_{k_*}=\|\sigma_{k_*+1}-\sigma_{k_*}\|_1\leq \epsilon_{\rm stop} .
  \label{eq:stopping}
\end{equation}
The contraction bound gives
\begin{equation}
  \|\sigma_{k_*}-\sigma_*\|_1
  \leq \frac{s_{k_*}}{1-\eta}
  \leq \frac{\epsilon_{\rm stop}}{1-\eta},
  \label{eq:correct-bound-current}
\end{equation}
or, if the reported estimate is the next iterate,
\begin{equation}
  \|\sigma_{k_*+1}-\sigma_*\|_1
  \leq \frac{\eta\,s_{k_*}}{1-\eta}
  \leq \frac{\eta}{1-\eta}\epsilon_{\rm stop} .
  \label{eq:correct-bound-next}
\end{equation}
Thus the factor \(\eta/(1-\eta)\) applies to the next iterate, whereas the current iterate uses the prefactor \(1/(1-\eta)\).  In the single-qubit toy example, using \(N_{\rm shots}=10^4\) per Pauli basis gives a step-size uncertainty of order \(3/\sqrt{N_{\rm shots}}\simeq0.03\).  A statistically resolved stopping threshold below this scale requires more shots or a different estimator.

\section{Numerical demonstrations}
\label{sec:numerics}

\subsection{Hardware parameter sweep}

We performed a coarse \(6\times8\) sweep over two-parameter perturbations to test whether the fidelity pattern predicted by a noiseless simulator survives hardware noise.  At each grid point \((\theta,\phi)\in[0,\pi]\times[0,2\pi]\), the local rotation
\begin{equation}
  V(\theta,\phi)=R_x(\theta)R_z(\phi)
  \label{eq:local-rotation}
\end{equation}
was inserted on register \(C\) before the scrambler.  State tomography on the recovered register \(Y\) was performed in the three Pauli bases.  The recovered-message fidelity was estimated as
\begin{equation}
  F_{\rm msg}(\theta,\phi)=\langle\psi_{\rm msg}|\rho_Y(\theta,\phi)|\psi_{\rm msg}\rangle .
\end{equation}

\begin{figure*}[t]
\centering
\includegraphics[width=\linewidth]{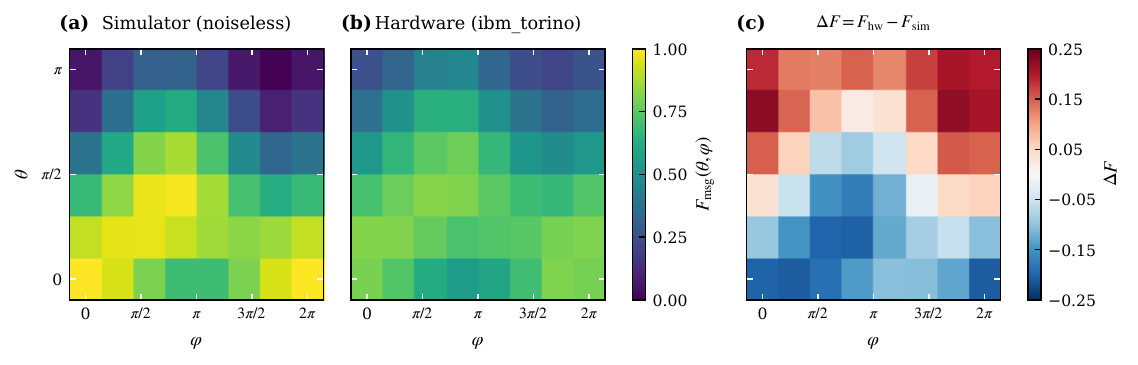}
\caption{Parameter sweep of the message-recovery fidelity \(F_{\rm msg}(\theta,\phi)\) over \(\theta\in[0,\pi]\), \(\phi\in[0,2\pi]\), using the local perturbation \(R_x(\theta)R_z(\phi)\) on register \(C\).  (a) Noiseless simulator prediction, with a high-fidelity region at small \(\theta\) and a low-fidelity region near \(\theta\simeq\pi\).  (b) Hardware data, which preserve the main \(\theta\)-dependent trend but compress the dynamic range toward mixed-state behavior.  (c) Pointwise discrepancy \(\Delta F=F_{\rm hw}-F_{\rm sim}\).  The sign pattern is consistent with decoherence pushing the recovered state toward a more mixed output.}
\label{fig:sweep}
\end{figure*}

The hardware preserves much of the rank ordering of grid points while compressing the dynamic range, as expected when noise drives the output toward a mixed state.  In the data set shown in Fig.~\ref{fig:sweep}, the Spearman rank correlation between simulator and hardware fidelities is
\begin{equation}
  r_S(F_{\rm hw},F_{\rm sim})=0.94,
  \label{eq:spearman}
\end{equation}
and the Pearson correlation is \(r_P=0.97\).  The high correlations show that the noiseless simulation captures the main ordering of perturbation angles on this device and calibration instance.

\subsection{Local perturbations on the early-radiation qubit}

We next study local unitary perturbations applied to the early-radiation qubit \(E\).  The message state is held fixed,
\begin{equation}
  |\psi\rangle=\cos(\theta_{\rm msg}/2)|0\rangle
  +e^{i\phi_{\rm msg}}\sin(\theta_{\rm msg}/2)|1\rangle,
  \label{eq:bloch-message}
\end{equation}
with \((\theta_{\rm msg},\phi_{\rm msg})\simeq(2.5,2.0)\).  Only the projection of its Bloch vector onto the perturbation axis,
\begin{equation}
  n_x=\sin\theta_{\rm msg}\cos\phi_{\rm msg},
\end{equation}
enters the analytic expressions below.

In Method I, \(R_X(\theta)\) is applied to \(E\) before scrambling.  The Bell-pair ricochet identity
\begin{equation}
  \bigl(R_X(\theta)\otimes I\bigr)|\Phi^+\rangle
  = \bigl(I\otimes R_X(\theta)\bigr)|\Phi^+\rangle
\end{equation}
moves the perturbation to the other end of the EPR resource.  In the ideal decoder branch, the recovered state is therefore \(R_X(\theta)|\psi\rangle\).  The fidelity with the original message is
\begin{equation}
  F_I(\theta)=|\langle\psi|R_X(\theta)|\psi\rangle|^2
  =\cos^2\frac{\theta}{2}+n_x^2\sin^2\frac{\theta}{2} .
  \label{eq:method-I}
\end{equation}
The value at \(\theta=\pi\) is the geometric overlap with a rotated state; the ideal decoder branch itself remains coherent.

\begin{figure*}[t]
\centering
\includegraphics[width=0.9\linewidth]{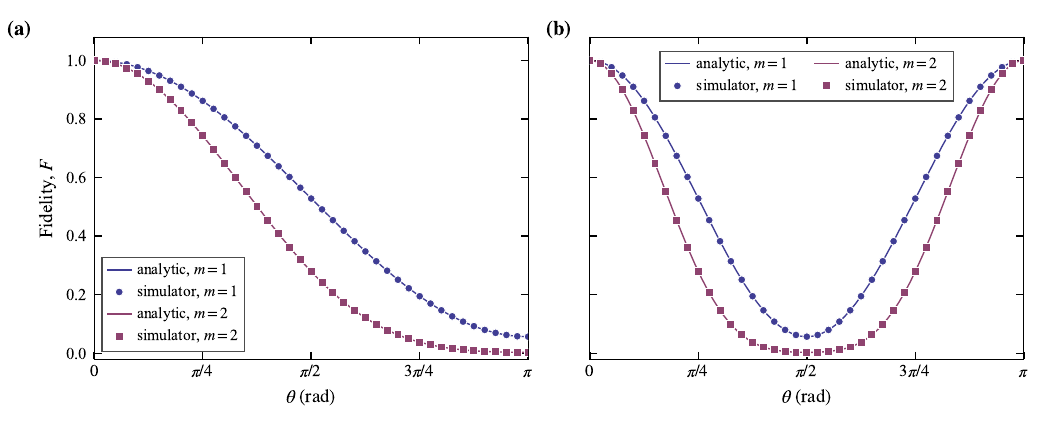}
\caption{Decoded fidelity \(F\) under local \(R_X(\theta)\) perturbations on the early-radiation qubit, for \(m=1\) and \(m=2\) parallel single-qubit decoder-emulation pipelines with identical message states.  (a) Method I: a single \(R_X(\theta)\) is applied to each \(E_i\) before scrambling.  The fidelity decreases from \(F(0)=1\) toward the geometric floor \(F_I^{(m)}(\pi)=n_x^{2m}\).  (b) Method II: echo-like sandwiches \(R_X(-\theta)UR_X(\theta)\) are inserted around the scrambler and decoder blocks.  The response is frequency-doubled, with a trough near \(\theta=\pi/2\) and revival at \(\theta=\pi\).  Solid curves are the analytic expressions in Eqs.~\eqref{eq:method-I}--\eqref{eq:method-II}; markers are noiseless statevector simulations.}
\label{fig:local-perturbations}
\end{figure*}

In Method II, the perturbation is inserted as conjugation sandwiches \(R_X(-\theta)UR_X(\theta)\) around the scrambler and decoder blocks.  In the post-selected sector this produces an effective \(R_X(2\theta)\) rotation of the recovered message, giving
\begin{equation}
\begin{aligned}
  F_{II}(\theta)
  &=|\langle\psi|R_X(2\theta)|\psi\rangle|^2\\
  &=\cos^2\theta+n_x^2\sin^2\theta .
\end{aligned}
  \label{eq:method-II}
\end{equation}
Equations~\eqref{eq:method-I} and \eqref{eq:method-II} have the common form
\begin{equation}
\begin{aligned}
  F_k(\theta)&=1-(1-n_x^2)\sin^2(\alpha_k\theta/2),\\
  \alpha_I&=1,\qquad \alpha_{II}=2 .
\end{aligned}
  \label{eq:unified-local}
\end{equation}
Thus the frequency doubling and revival in Method II come from the sandwich construction.  Statevector simulation agrees with the analytic expressions to machine precision, as shown in Fig.~\ref{fig:local-perturbations}.  Method I measures uncompensated unitary overlap loss, while Method II measures an echo-like response of the decoder branch.

\begin{figure*}[t]
\centering
\includegraphics[width=\linewidth]{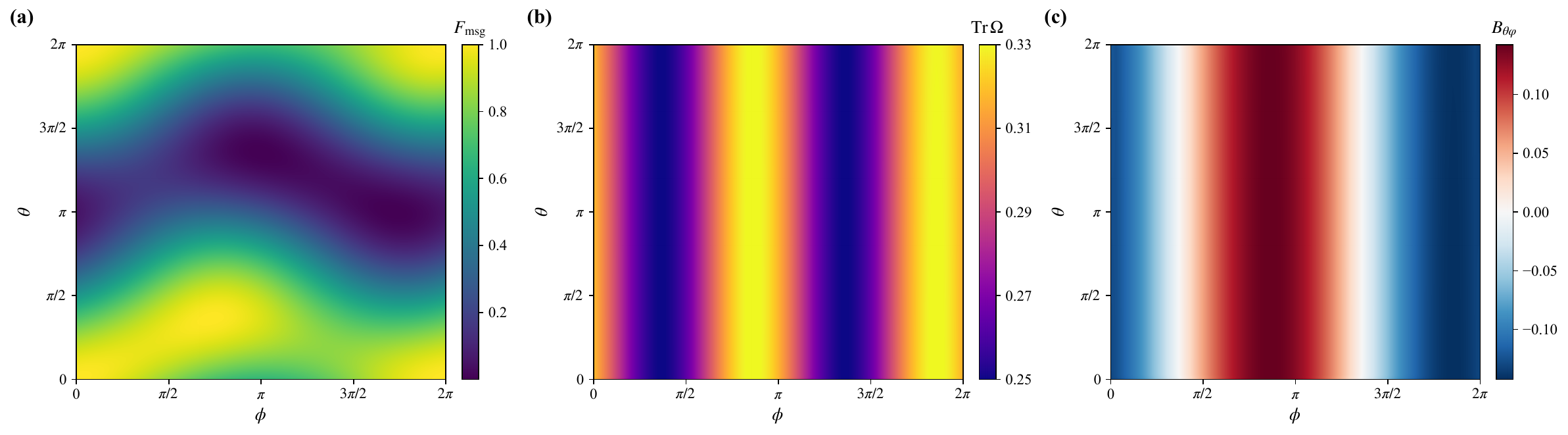}
\caption{Two-parameter perturbation landscape for \(R_x(\theta)R_z(\phi)\) applied to register \(C\), with the message state held fixed and the full circuit evaluated by exact statevector simulation.  (a) Recovered-message fidelity \(F_{\rm msg}(\theta,\phi)\), the operational decoder metric.  (b) Trace of the QGT metric part, \(\mathrm{Tr}\,\Omega(\theta,\phi)\), used here as a coordinate-fixed susceptibility of the global pre-measurement state.  (c) Berry-curvature-like component \(B_{\theta\phi}=\mathrm{Im}\,Q_{\theta\phi}\) in the convention of Eq.~\eqref{eq:qgt-parts}.  The metric and curvature panels show mainly \(\phi\)-dependent striping, whereas the recovered-message fidelity varies most strongly with \(\theta\).}
\label{fig:qgt}
\end{figure*}

\subsection{Two-parameter perturbations and the quantum geometric tensor}

We also use quantum-geometric diagnostics for the global pre-measurement output state.  The local perturbation
\begin{equation}
  V(\theta,\phi)=R_x(\theta)R_z(\phi)
  \label{eq:vtheta}
\end{equation}
is inserted on register \(C\), so the effective scrambler is
\begin{equation}
  U_{\rm scr}(\theta,\phi)=U\bigl(V(\theta,\phi)_C\otimes I_{ER}\bigr) .
  \label{eq:perturbed-scrambler}
\end{equation}
For each parameter pair, exact statevector simulation gives a global pure state \(|\Psi(\theta,\phi)\rangle\in(\mathbb{C}^2)^{\otimes7}\).  The operational recovered-message fidelity is computed from the reduced state on \(Y\):
\begin{equation}
  F_{\rm msg}(\theta,\phi)=
  \langle\psi_{\rm msg}|\rho_Y(\theta,\phi)|\psi_{\rm msg}\rangle .
  \label{eq:fmsg-qgt}
\end{equation}

For the global pure-state manifold, the quantum geometric tensor is
\begin{equation}
\begin{aligned}
  Q_{\mu\nu}
  &=\langle\partial_\mu\Psi|\partial_\nu\Psi\rangle
   -\langle\partial_\mu\Psi|\Psi\rangle
    \langle\Psi|\partial_\nu\Psi\rangle,\\
  &\hspace{6em}\lambda^\mu\in\{\theta,\phi\}.
\end{aligned}
  \label{eq:qgt-correct}
\end{equation}
We define
\begin{equation}
\begin{aligned}
  \Omega_{\mu\nu}(\theta,\phi)&:=\mathrm{Re}\,Q_{\mu\nu}(\theta,\phi),\\
  B_{\theta\phi}(\theta,\phi)&:=\mathrm{Im}\,Q_{\theta\phi}(\theta,\phi).
\end{aligned}
\label{eq:qgt-parts}
\end{equation}
Here \(\Omega_{\mu\nu}\) is the Fubini--Study metric in the chosen coordinates, while \(B_{\theta\phi}\) is a Berry-curvature-like component.  Some conventions define the Berry curvature with an additional factor or sign; Eq.~\eqref{eq:qgt-parts} fixes the convention used here.

For a small displacement \(\dd\lambda=(\dd\theta,\dd\phi)\),
\begin{equation}
\begin{aligned}
  &1-|\langle\Psi(\lambda)|\Psi(\lambda+\dd\lambda)\rangle|^2\\
  &\qquad=\Omega_{\mu\nu}(\lambda)\,\dd\lambda^\mu\dd\lambda^\nu
  +O(\|\dd\lambda\|^3).
\end{aligned}
  \label{eq:fs-distance}
\end{equation}
We summarize local sensitivity in the fixed coordinate chart by
\begin{equation}
  \tr\Omega(\theta,
  \phi)=\Omega_{\theta\theta}(\theta,\phi)+\Omega_{\phi\phi}(\theta,\phi) .
  \label{eq:metric-trace}
\end{equation}
This scalar is used as a coordinate-fixed susceptibility in the chosen parametrization.

Fig. ~\ref{fig:qgt} compares \(F_{\rm msg}\), \(\tr\Omega\), and \(B_{\theta\phi}\).  The post-selection probability is comparatively featureless over the same parameter grid, while the recovered-message fidelity has a pronounced \(\theta\)-dependent structure.  The metric trace and the Berry-curvature-like component mainly show \(\phi\)-dependent striping, as derived in Appendix~\ref{app:qgt-echo}.  These quantities therefore diagnose different aspects of the circuit response: \(F_{\rm msg}\) is an operational recovery metric, whereas QGT components describe the sensitivity of the global state before measurement.

\section{Conclusion}

We studied a finite-dimensional circuit toy model in which a Yoshida--Kitaev-type decoder is placed inside a Deutsch fixed-point loop.  The ideal result is a register-routing statement.  The incoming CTC state is moved to an idle dump register, the active branch recovers the prepared message on \(Y\), and the final SWAP writes that message into \(C\).  Under these conditions the induced Deutsch map on \(C\) is the replacement channel \(\sigma\mapsto\rho_M\), and the unique fixed point is \(\sigma_* = \rho_M\).

We then implemented the post-selected decoder branch with gate-model hardware, measuring recovery fidelity and sampling overhead.  We also formulated a classical-feedback iteration for fixed points of an estimated CPTP map and derived its contraction-based stopping bound.  Together, these results connect the ideal fixed-point theorem to experimentally measurable decoder observables.

The single-qubit hardware data show that the main fidelity landscape predicted by simulation persists under realistic noise, with the expected compression of the dynamic range.  The QGT and Loschmidt-echo measurements add a complementary view of global-state sensitivity and separate noise contributions from post-selection-rate effects.  The main scaling challenge is the exponential post-selection or amplitude-amplification overhead with message size; a full fixed-point process certificate would additionally require direct tomography of the CTC-input map.

\begin{acknowledgments}
 S.N.M would like to thank Maganti Gayathri for helping in creating Figure 1. This work was supported by the NSF under Grant No. OSI-2328774.
\end{acknowledgments}

\section*{Code availability}
All simulation code, perturbation-sweep data, and figure-generation scripts that support the findings of this study are openly available at Ref.~\cite{Morapakula2026}.

\section*{Author contributions}

S.N.M. developed the circuit constructions, performed the simulations and hardware experiments, analyzed the data, prepared the figures, and wrote the initial manuscript. K.I. supervised the project, contributed to the theoretical formulation and interpretation, and revised the manuscript.

A large language model was used for language editing, LaTeX reformatting. All scientific arguments, derivations, citations, code, and numerical outputs were checked by the authors, who take full responsibility for the manuscript.

\appendix

\section{Decoder-process validation and relation to the approximate fixed-point bound}
\label{app:qpt}

We verified the single-qubit deterministic decoder emulator numerically using the AerSimulator by preparing ten message states \(\rho_M\), including \(|0\rangle\), \(|1\rangle\), \(|+\rangle\), \(|+i\rangle\), the hardware message state in Eq.~\eqref{eq:message-state}, and five random parameter choices.  For the ideal compressed decoder circuit, the reduced output on the recovered register satisfies \(F(\rho_Y,\rho_M)=1\) and \(\|\rho_Y-\rho_M\|_F=0\) to numerical precision.  This validates the message-recovery action of the compressed decoder branch; the arbitrary-input Deutsch fixed-point statement is the register-separated theorem of Sec.~\ref{sec:deutsch-loop}.

The QPT procedure below varies the prepared message and reconstructs an effective message-to-output process.  This process characterizes the measured decoder branch, whereas Lemma~II.2 concerns the CTC-input map \(\Phi_{\rho_{\mathrm{CR}}}\) at fixed \(\rho_M\).

Let \(\Phi_{M\to C}\) denote the effective message-to-output process reconstructed from hardware data.  We characterize its affine action on the Bloch ball,
\begin{equation}
  \bm r_{\rm out}=\bm t+T\bm r_{\rm in},
  \label{eq:affine-map}
\end{equation}
where \(T\in\mathbb{R}^{3\times3}\) and \(\bm t\in\mathbb{R}^3\).  Four linearly independent input states, \(|0\rangle\), \(|1\rangle\), \(|+\rangle\), and \(|+i\rangle\), were prepared, and the output was measured in the three Pauli bases.  Linear inversion followed by eigenvalue clipping gives a physical output density matrix for each probe, and \((T,\bm t)\) is obtained from the resulting linear system.  Because this reconstruction does not impose a global CPTP constraint on the fitted process, we report Bloch-map quantities directly.

For the deviation from the identity process, \(\mathcal{E}=\Phi_{M\to C}-\mathrm{id}\), the Bloch action is
\begin{equation}
  \bm r\mapsto \bm t+(T-I_3)\bm r .
\end{equation}
We report the single-qubit Bloch-ball diagnostic
\begin{equation}
  \Delta_{\rm Bloch}:=|\bm t|+\sigma_{\max}(T-I_3),
  \label{eq:bloch-diagnostic}
\end{equation}
which upper-bounds the observed output Bloch-vector deviation for unentangled single-qubit inputs.  A diamond-norm value for this channel would require a complete-bounded-norm calculation, for example by semidefinite programming.  We therefore report Eq.~\eqref{eq:bloch-diagnostic} as a Bloch-ball process diagnostic.

The reconstructed process is close to the identity on the tested message-to-output channel.  The contraction matrix has diagonal elements approximately \((T_{xx},T_{yy},T_{zz})=(0.99,1.00,1.00)\), off-diagonal elements below \(0.02\), and \(|\bm t|=0.008\).  Table~\ref{tab:qpt} summarizes the resulting diagnostics.

\begin{table}[b]
\centering
\caption{QPT diagnostics for the message-to-output process \(\Phi_{M\to C}\) on \backend{ibm_kingston}. The quantity \(\Delta_{\rm Bloch}\) is defined in Eq.~\eqref{eq:bloch-diagnostic}.}
\label{tab:qpt}
\small
\begin{tabular}{@{}lcc@{}}
\toprule
Quantity & Value & 95\% CI \\
\midrule
\(\Delta_{\rm Bloch}\) & 0.03 & [0.02, 0.07] \\
\bottomrule
\end{tabular}
\end{table}

These data show that the measured message-to-output process is close to the ideal decoder process on the tested backend.  The corresponding CTC-input certificate would be obtained by reconstructing \(\Phi_{\rho_{\mathrm{CR}}}\) at fixed \(\rho_M\) and bounding \(\|\Phi_{\rho_{\mathrm{CR}}}-\mathcal{R}_{\rho_M}\|_\diamond\).

\section{Deterministic decoder details}
\label{app:det-decoder}

\subsection{Scrambling unitary and Grover reflection}

For the emulations, we use the scrambler
\begin{equation}
\begingroup
\setlength{\arraycolsep}{2.5pt}
U=\frac{1}{2\sqrt{2}}
\begin{pmatrix}
 1& 1& 1&-1& 1&-1&-1&-1\\
 1&-1& 1& 1& 1& 1&-1& 1\\
 1& 1&-1& 1& 1&-1& 1& 1\\
-1& 1& 1& 1&-1&-1&-1& 1\\
 1& 1& 1&-1&-1& 1& 1& 1\\
-1& 1&-1&-1& 1& 1&-1& 1\\
-1&-1& 1&-1& 1&-1& 1& 1\\
-1& 1& 1& 1& 1& 1& 1&-1
\end{pmatrix}.
\endgroup
\label{eq:scrambler}
\end{equation}
It has operator-spreading properties of the kind used in scrambling experiments \cite{Landsman2019}.

Following Yoshida and Kitaev \cite{YoshidaKitaev2017}, the active decoder is built from a Grover iterate
\begin{equation}
  W \equiv \widetilde{W}_A W_D,
  \label{eq:grover-iterate}
\end{equation}
where \(W_D=I-2P_D\) reflects about the Bell projector \(P_D=|\Phi^+\rangle\langle\Phi^+|\) on \((R,G)\), and
\begin{equation}
  \widetilde{W}_A=(I\otimes U^*)(2P_A-I)(I\otimes U^T)
\end{equation}
reflects about the decoded target subspace on \((A,Y)\) in the conjugated form used by Ref.~\cite{YoshidaKitaev2017}.  For \(d_A=d_R=2^m\), define the amplitude-amplification angle \(\vartheta_m=\arcsin(2^{-m})\).  After \(k\) Grover iterates, the success probability is \(\sin^2[(2k+1)\vartheta_m]\), so the optimal integer count is
\begin{equation}
\begin{aligned}
  N_W(m)&=\operatorname*{arg\,min}_{k\in\mathbb{Z}_{\geq0}}
  \left|(2k+1)\vartheta_m-\frac{\pi}{2}\right|,\\
  N_W(m)&=\frac{\pi}{4}2^m+O(1).
\end{aligned}
  \label{eq:grover-count}
\end{equation}
For \(m=1\), \(\vartheta_1=\pi/6\) and \(N_W=1\).

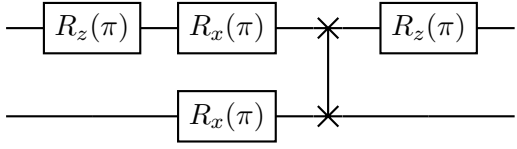
\begin{figure}[t]
    \centering
    \begin{quantikz}
\lstick{} & \gate{R_z(\pi)} & \gate{R_x(\pi)} & \swap{1} & \gate{R_z(\pi)} & \qw \\
\lstick{} & \qw             & \gate{R_x(\pi)} & \targX{} & \qw             & \qw
\end{quantikz}
    \caption{Gate representation of the Bell-state Grover reflection \(W_D\).}
\label{fig:grover-reflection}
\end{figure}

In the computational basis of \((R,G)\), the Bell reflection used here is
\begin{equation}
W_D=\begin{pmatrix}
0&0&0&-1\\
0&1&0&0\\
0&0&1&0\\
-1&0&0&0
\end{pmatrix},
\label{eq:WD}
\end{equation}
which sends \(|\Phi^+\rangle\mapsto-|\Phi^+\rangle\) and acts as the identity on the orthogonal Bell states.  A gate representation of this reflection is shown in Fig.~\ref{fig:grover-reflection}.

In the hardware-emulator circuit, the decoder block after the scrambler contains the Bell-reflection pattern used in the Grover iterate and a final Bell-basis projection.  The final reflection is redundant on the post-selected \(|\Phi^+\rangle_{RG}\) branch and is retained only as part of the chosen gate-level layout.  Operationally, the post-selected decoder is equivalent to the single Grover iterate followed by the Bell projection.

\subsection{Resource estimates}

For the fixed-iterate circuit used in the single-qubit demonstration, the pre-transpilation two-qubit gate count scales linearly with \(m\).  The full deterministic Yoshida--Kitaev decoder also includes the number of Grover iterates in Eq.~\eqref{eq:grover-count}.  Table~\ref{tab:gate-count} reports the fixed-iterate counts, while Table~\ref{tab:grover-scaling} shows the exponential growth of the amplitude-amplification depth.

\begin{table}[t]
\centering
\caption{Fixed-iterate decoder-emulation resource counts before and after transpilation. The full deterministic cost also includes the Grover-iteration scaling in Table~\ref{tab:grover-scaling}.}
\label{tab:gate-count}
\small
\setlength{\tabcolsep}{4pt}
\begin{tabular}{@{}cccc@{}}
\toprule
\(m\) & Logical qubits & \(N_{2Q}^{\rm pre}\) & \(N_{2Q}^{\rm post}\) \\
\midrule
1 & 7  & 33 & 106 \\
2 & 14 & 66 & 209 \\
3 & 21 & 99 & 318 \\
\bottomrule
\end{tabular}
\end{table}

\begin{table*}[t]
\centering
\caption{Amplitude-amplification iteration counts from \(\vartheta_m=\arcsin(2^{-m})\). The second column is the leading asymptotic estimate. The ratios approach two, consistent with \(N_W(m)=\Theta(2^m)\).}
\label{tab:grover-scaling}
\small
\setlength{\tabcolsep}{10pt}
\begin{tabular}{@{}cccc@{}}
\toprule
\(m\) & Asymptotic \(\pi 2^m/4\) & Optimal \(N_W(m)\) & Ratio \(N_W(m)/N_W(m-1)\) \\
\midrule
1 & 1.571 & 1   & -- \\
2 & 3.142 & 3   & 3.00 \\
3 & 6.283 & 6   & 2.00 \\
4 & 12.566 & 12 & 2.00 \\
5 & 25.133 & 25 & 2.08 \\
6 & 50.265 & 50 & 2.00 \\
7 & 100.531 & 100 & 2.00 \\
8 & 201.062 & 201 & 2.01 \\
9 & 402.124 & 402 & 2.00 \\
10 & 804.248 & 804 & 2.00 \\
\bottomrule
\end{tabular}
\end{table*}

The implemented single-iterate circuit matches the deterministic protocol for \(m=1\).  For \(m\geq2\), the full deterministic decoder includes additional amplitude-amplification steps.  At \(m=10\), the count exceeds 800 Grover iterates, setting the near-term hardware scale of the deterministic construction.

\section{Probabilistic decoder data}
\label{app:prob-decoder}

The post-selected implementation runs the linear circuit and retains shots with the selected Bell outcome \(|\Phi^+\rangle\).  Conditioned on success, the retained branch matches the ideal decoder action for the initialized circuit and prepares \(Y\) close to the message state.  This section reports the decoder-recovery data underlying the hardware results.

For each hardware data set, the relevant reproducibility metadata are the IBM job identifiers, execution dates, Qiskit version, transpiler seed, backend name, and calibration timestamp.  The backend names below identify the data sets and routing/calibration models used in the present analysis.

\begin{table*}[t]
\centering
\caption{Representative transpilation results across IBM Quantum backends at optimization level 3.}
\label{tab:transpilation}
\small
\setlength{\tabcolsep}{12pt}
\begin{tabular}{@{}lccc@{}}
\toprule
Backend & Qubits & Transpiled depth & Two-qubit gates \\
\midrule
\backend{ibm_torino} & 133 & 87 & 32 \\
\backend{ibm_fez} & 156 & 87 & 35 \\
\backend{ibm_marrakesh} & 156 & 87 & 35 \\
\backend{ibm_kingston} & 156 & 87 & 35 \\
\bottomrule
\end{tabular}
\end{table*}

\begin{table}[t]
\centering
\caption{Calibration snapshot for the seven physical qubits used by the routed \backend{ibm_torino} circuit.}
\label{tab:calibration}
\footnotesize
\setlength{\tabcolsep}{4pt}
\begin{tabular}{@{}lcc@{}}
\toprule
Qubit & Register & Readout error \\
\midrule
q45 & \(C\) & 0.79\% \\
q46 & \(E\) & 4.54\% \\
q55 & \(R\) & 0.85\% \\
q65 & \(G\) & 1.16\% \\
q66 & \(M\) & 0.63\% \\
q67 & \(A\) & 5.54\% \\
q68 & \(Y\) & 0.77\% \\
\midrule
Average & -- & 2.04\% \\
\bottomrule
\end{tabular}

\vspace{0.7em}
\begin{tabular}{@{}lc@{}}
\toprule
Edge & CZ error \\
\midrule
q45--q46 & 0.27\% \\
q46--q55 & 0.16\% \\
q55--q65 & 0.21\% \\
q65--q66 & 0.31\% \\
q66--q67 & 0.30\% \\
q67--q68 & 0.22\% \\
\midrule
Average & 0.25\% \\
\bottomrule
\end{tabular}
\end{table}

\begin{figure*}[t]
\centering
\includegraphics[width=\linewidth]{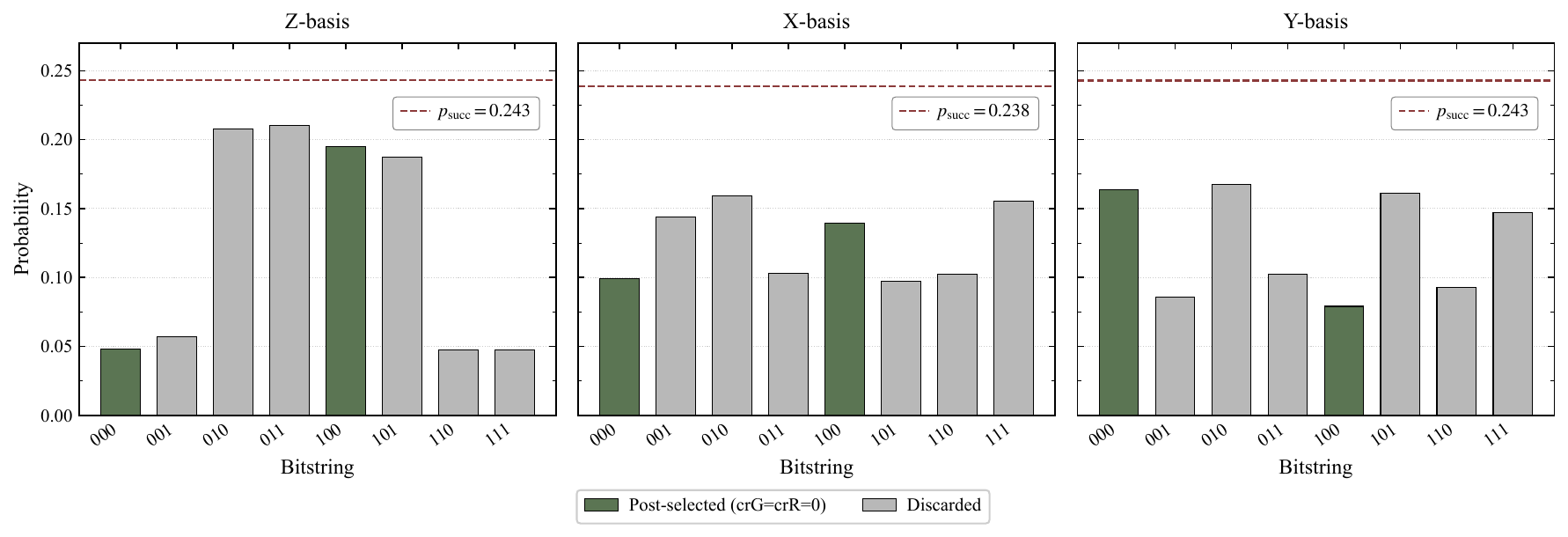}
\caption{Raw measurement histograms for the \(Z\), \(X\), and \(Y\) Pauli-basis circuits.  Green bars denote the retained outcomes with \((r,g)=(0,0)\), corresponding to the selected \(|\Phi^+\rangle\) Bell projection.  Gray bars are discarded outcomes.  The red dashed line marks \(p_{\rm succ}\) for each basis.  The post-selected fraction is stable across the three basis circuits, supporting the use of a common post-selection probability in the reconstruction.}
\label{fig:histograms}
\end{figure*}

\begin{figure}[t]
\centering
\includegraphics[width=0.8\linewidth]{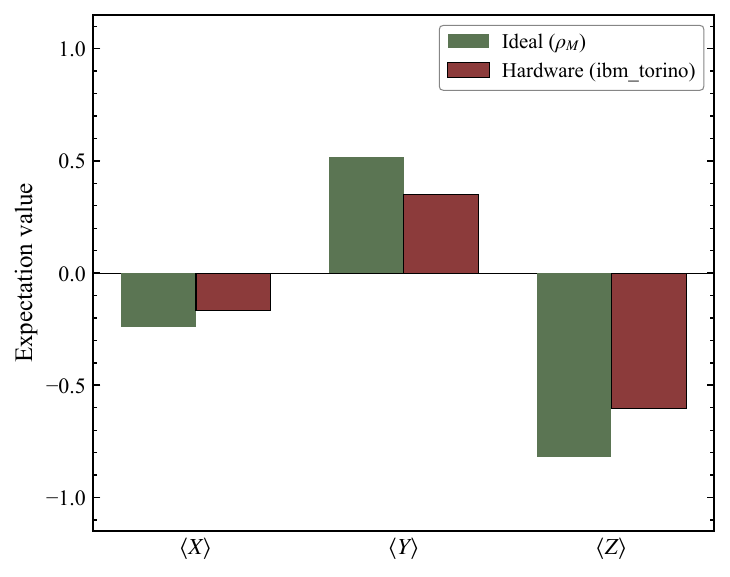}
\caption{Pauli expectation values \(\langle X\rangle\), \(\langle Y\rangle\), and \(\langle Z\rangle\) on the recovered register after post-selection.  The ideal values are those of the prepared message state \(\rho_M\).  The hardware values are attenuated relative to the ideal Bloch vector but preserve the sign pattern, consistent with decoder recovery followed by noise.}
\label{fig:pauli-values}
\end{figure}

The retained shots in each Pauli basis give estimates of \(\langle X\rangle\), \(\langle Y\rangle\), and \(\langle Z\rangle\) on register \(Y\).  We reconstruct
\begin{equation}
  \rho_Y=\frac{1}{2}\left(I+\langle X\rangle\sigma_x+\langle Y\rangle\sigma_y+\langle Z\rangle\sigma_z\right),
  \label{eq:linear-inversion}
\end{equation}
followed by projection to the nearest positive semidefinite unit-trace matrix.  The raw retained and discarded counts are shown in Fig.~\ref{fig:histograms}, and the reconstructed Pauli components are shown in Fig.~\ref{fig:pauli-values}.  The recovered-message fidelity is
\begin{equation}
  F(\rho_Y,\rho_M)=\langle\psi_M|\rho_Y|\psi_M\rangle .
  \label{eq:recovered-fidelity}
\end{equation}

\begin{figure*}[t]
\centering
\includegraphics[width=\linewidth]{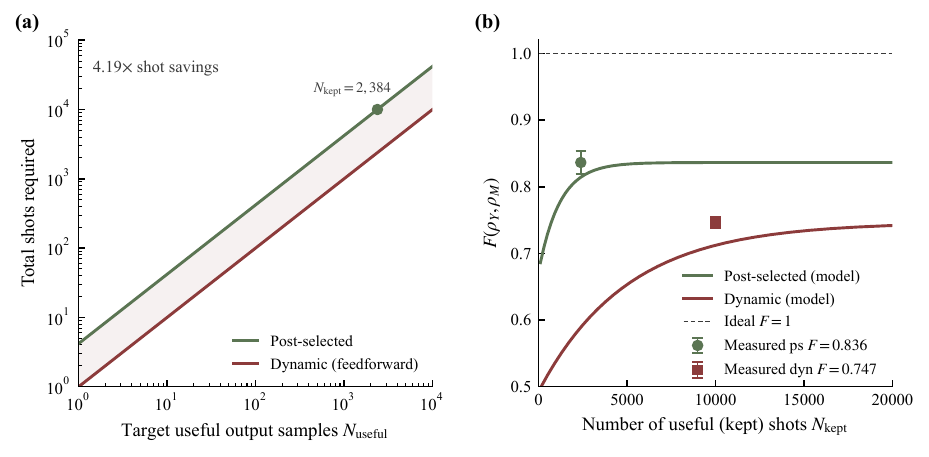}
\caption{Fidelity--shot-cost tradeoff between strict post-selection and dynamic feedforward.  (a) Total shots required to obtain a target number of useful samples; dynamic feedforward retains every shot and gives about a fourfold shot saving at the displayed operating point.  (b) Recovered-message fidelity as a function of useful kept shots.  The dynamic protocol yields more samples and narrower statistical intervals at fixed total shots, while the strict post-selected branch gives the higher conditional fidelity in this data set.}
\label{fig:shot-cost}
\end{figure*}

A hardware-native alternative replaces post-selection with conditional Pauli byproduct correction using dynamic circuits.  After the Bell measurement on \((R,G)\), a classical feedforward signal applies
\begin{equation}
(r,g)\mapsto
\begin{cases}
I,  & r=0,\ g=0,\\
Z,  & r=1,\ g=0,\\
X,  & r=0,\ g=1,\\
XZ, & r=1,\ g=1.
\end{cases}
\label{eq:feedforward}
\end{equation}
Every shot is retained, so the shot cost is reduced by roughly \(1/p_{\rm succ}\).  In the data set used here, the dynamic protocol gives about a factor of four more useful samples at fixed total shots, but with a lower fidelity point estimate than strict post-selection. 
Fig.~\ref{fig:shot-cost} summarizes this fidelity--shot-cost tradeoff.

\subsection{Error mitigation and noisy-simulator validation}

We compared three post-selected runs: an unmitigated baseline, a dynamical-decoupling run with XX sequences, and a zero-noise extrapolation (ZNE) run using gate folding at noise amplification factors \(\lambda=1,3,5\).  The values in Table~\ref{tab:mitigation} use a single master-data convention for the post-selected baseline and post-selection probability.

\begin{table*}[t]
\centering
\caption{Error-mitigation summary for the post-selected decoder-emulation data set. Values are rounded consistently with the master data.}
\label{tab:mitigation}
\small
\setlength{\tabcolsep}{12pt}
\begin{tabular}{@{}lccc@{}}
\toprule
Condition & \(F(\rho_Y,\rho_M)\) & \(\Delta F\) & \(p_{\rm succ}\) \\
\midrule
Unmitigated & 0.839 & -- & 0.235 \\
DD (XX sequences) & 0.752 & \(-0.087\) & 0.241 \\
ZNE (\(\lambda=1,3,5\)) & 0.842 & \(+0.003\) & 0.238 \\
Ideal noiseless & 1.000 & -- & 0.250 \\
\bottomrule
\end{tabular}
\end{table*}

The observed degradation under dynamical decoupling is consistent with the additional pulse overhead increasing the effective circuit depth on this calibration.  ZNE gives a marginal improvement, as expected for a circuit whose base two-qubit error rates are already small and whose loss also includes readout and routing-dependent contributions.  These mitigation results are device and calibration-specific.

We also compared the hardware data with simulator models built from the calibration snapshot.  The key distinction is whether the noise model is applied to the shallow logical seven-qubit circuit or to the fully routed heavy-hex circuit.  Local models without routing overestimate the recovered-message fidelity.  The routing-aware model, generated from the backend coupling map and calibration data, gives a much closer result.  Table~\ref{tab:noisy-simulator} summarizes the comparison.

\begin{table}[t]
\centering
\caption{Noisy-simulator validation for the post-selected data set. The routing-aware model reduces the gap because noise is applied after heavy-hex routing and gate decomposition. Here \(\Delta F=F_{\rm sim}-F_{\rm hw}\).}
\label{tab:noisy-simulator}
\scriptsize
\setlength{\tabcolsep}{3pt}
\begin{tabular}{@{}lccc@{}}
\toprule
Model & \(F_{\rm sim}\) & \(\Delta F\) & \(p_{\rm succ}^{\rm sim}\) \\
\midrule
NM0: Noiseless & 0.997 & +0.140 & 0.250 \\
NM1: Depolarizing only & 0.995 & +0.138 & 0.250 \\
NM2: Depolarizing + \(T_1/T_2\) & 0.995 & +0.138 & 0.250 \\
NM3: Scaled local model & 0.998 & +0.141 & 0.250 \\
NM4: Routing-aware backend model & 0.893 & +0.036 & 0.243 \\
Hardware & 0.857 & -- & 0.243 \\
\bottomrule
\end{tabular}
\end{table}

A simple scale check using the mean CZ error rate illustrates the issue.  Applying noise to a shallow local circuit with 16 two-qubit gates gives roughly
\begin{equation}
  (1-\bar\epsilon_{\rm CZ})^{16}\simeq0.96,
\end{equation}
whereas the routed hardware circuit has 32 two-qubit gates and depth 87, giving
\begin{equation}
  (1-\bar\epsilon_{\rm CZ})^{32}\simeq0.92 .
\end{equation}
This estimate is a scale check rather than a direct map from depolarizing error to message infidelity.  It explains why local noise models are too optimistic.  NM4 captures the dominant routing overhead, and the remaining gap is consistent with coherent miscalibration, residual ZZ crosstalk, calibration drift, and readout effects beyond the calibration-derived model.

\newpage
\section{Hardware-accessible QGT proxy via Loschmidt echo}
\label{app:qgt-echo}

Full tomography of the global seven-qubit state \(|\Psi(\theta,\phi)\rangle\) is impractical on hardware.  We instead estimate diagonal Fubini--Study metric components through a Loschmidt echo~\cite{Wisniacki2012}.  For a small step \(\delta\),
\begin{equation}
  1-|\langle\Psi(\theta,\phi)|\Psi(\theta+\delta,\phi)\rangle|^2
  =\Omega_{\theta\theta}(\theta,\phi)\delta^2+O(\delta^4) .
  \label{eq:loschmidt-metric}
\end{equation}
The overlap is measured by the echo circuit
\begin{equation}
  \mathcal{E}_\theta(\delta)=U^\dagger(\theta+\delta,\phi)U(\theta,
  \phi),
  \label{eq:theta-echo}
\end{equation}
applied to \(|0\rangle^{\otimes7}\).  The all-zeros probability satisfies
\begin{equation}
\begin{aligned}
  P_{00}(\delta)
  &=|\langle0\cdots0|\mathcal{E}_\theta(\delta)|0\cdots0\rangle|^2\\
  &=|\langle\Psi(\theta,\phi)|\Psi(\theta+\delta,\phi)\rangle|^2 .
\end{aligned}
  \label{eq:p00}
\end{equation}
Thus \(\Omega_{\theta\theta}\simeq(1-P_{00})/\delta^2\) for sufficiently small \(\delta\), or more accurately the finite-step expression below can be fitted.

The gate-ordering structure gives an exact simplification.  Write
\begin{equation}
  U(\theta,\phi)=U_2 R_x(\theta)R_z(\phi)U_1 .
  \label{eq:u-factorized}
\end{equation}
Then
\begin{align}
  \mathcal{E}_\theta(\delta)
  &=U_1^\dagger R_z(-\phi)R_x[-(\theta+
  \delta)]R_x(\theta)R_z(\phi)U_1 \\
  &=U_1^\dagger R_z(-\phi)R_x(-\delta)R_z(\phi)U_1 .
  \label{eq:theta-cancellation}
\end{align}
The \(\theta\) rotations cancel, while the \(\phi\) rotations sandwich the residual \(R_x(-\delta)\).  If the reduced state of register \(C\) in \(|\chi\rangle=U_1|0\rangle^{\otimes7}\) is
\begin{equation}
  \rho_C=\frac{1}{2}(I+r_x\sigma_x+r_y\sigma_y+r_z\sigma_z),
\end{equation}
then direct evaluation gives
\begin{equation}
  P_{00}(\delta;\phi)=1-4\Omega_{\theta\theta}(\phi)\sin^2\frac{\delta}{2},
  \label{eq:p00-exact}
\end{equation}
with
\begin{equation}
  \Omega_{\theta\theta}(\phi)=\frac{1-(r_x\cos\phi-r_y\sin\phi)^2}{4} .
  \label{eq:omega-theta}
\end{equation}

The companion \(\phi\)-direction echo is
\begin{equation}
  \mathcal{E}_\phi(\delta_\phi)=U^\dagger(\theta,\phi+
  \delta_\phi)U(\theta,\phi)=U_1^\dagger R_z(-\delta_\phi)U_1,
  \label{eq:phi-echo}
\end{equation}
so both base-point parameters cancel.  Hence
\begin{equation}
  \Omega_{\phi\phi}=\frac{1-r_z^2}{4} .
  \label{eq:omega-phi}
\end{equation}
For the Bloch vector used in the circuit, \((r_x,r_y,r_z)=(-0.23,+0.51,-0.82)\), this gives \(\Omega_{\phi\phi}\simeq0.08\).  The metric trace therefore depends on \(\phi\) but not on \(\theta\) in the ideal circuit.

\begin{figure*}[t]
\centering
\includegraphics[width=\linewidth]{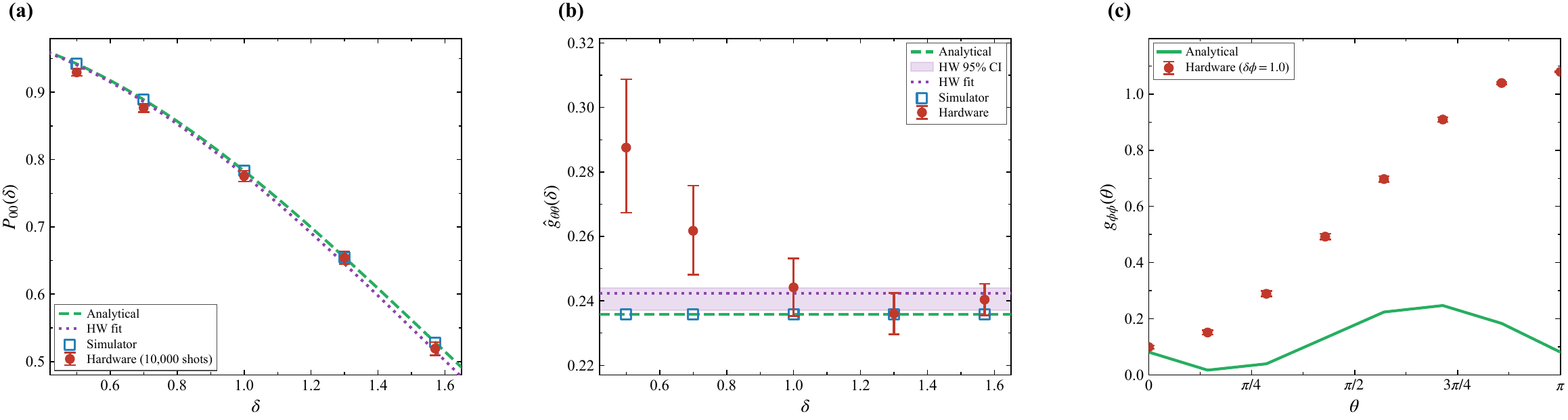}
\caption{Loschmidt-echo diagnostics.  (a) All-zeros probability \(P_{00}(\delta)\) for the \(\theta\)-echo at \(\phi=0\), compared with the noiseless simulator, the analytical curve, and a hardware fit.  (b) Per-step estimate \((1-P_{00})/[4\sin^2(\delta/2)]\), which is constant in the ideal circuit according to Eq.~\eqref{eq:p00-exact}.  (c) Hardware estimate of the \(\phi\)-direction proxy \(\Omega_{\phi\phi}(\theta)\), compared with the ideal constant prediction of Eq.~\eqref{eq:omega-phi}.  The growth at large \(\theta\) reflects noise-induced echo suppression in the transpiled circuit.}
\label{fig:loschmidt}
\end{figure*}

The same factorization gives the Berry-curvature-like component in the convention of Eq.~\eqref{eq:qgt-parts}.  Using \(\partial_\theta R_x(\theta)=-(i/2)\sigma_xR_x(\theta)\) and \(\partial_\phi R_z(\phi)=-(i/2)\sigma_zR_z(\phi)\), one obtains
\begin{equation}
  B_{\theta\phi}(\phi)=-\frac{1}{4}(r_x\sin\phi+r_y\cos\phi).
  \label{eq:BthetaPhi}
\end{equation}
This expression is independent of \(\theta\) for the same gate-ordering reason as Eq.~\eqref{eq:omega-theta}.  It matches the \(\phi\)-striped structure in Fig.~\ref{fig:qgt}.  The corresponding Loschmidt-echo measurements are shown in Fig.~\ref{fig:loschmidt}.

The \(\theta\)-echo was run at \(\phi=0\) with five step sizes, \(\delta\in\{0.5,0.7,1.0,1.3,1.5\}\), using \(N_{\rm shots}=10^4\) shots per circuit.  Fitting Eq.~\eqref{eq:p00-exact} gives
\begin{equation}
  \Omega_{\theta\theta}^{\rm hw}(\phi=0)=0.242\pm0.002,
  \label{eq:omega-hw}
\end{equation}
The fitted value is 0.006 above the ideal result \(\Omega_{\theta\theta}^{\rm exact}(\phi=0)=0.236\), a shift larger than the quoted statistical uncertainty.  A separate identity echo gave \(P_{00}^{\rm id}=0.9816\), identifying a 1.84\% systematic noise floor with the same upward effect on the extracted metric.

Equation~\eqref{eq:omega-phi} predicts that \(\Omega_{\phi\phi}\) is constant.  Hardware estimates with \(\delta_\phi=1.0\) agree near \(\theta=0\) but rise strongly at larger \(\theta\).  The rise reflects noise-induced suppression of \(P_{00}\) in the depth-87 transpiled echo circuit: when the noisy overlap is fit to the ideal expression, the noise contribution appears as an inflated effective metric.  The predicted constancy of \(\Omega_{\phi\phi}\) therefore makes this echo a sensitive hardware-noise probe.


\begin{thebibliography}{99}

\bibitem{Godel1949}
K. Godel, ``An Example of a New Type of Cosmological Solutions of Einstein's Field Equations of Gravitation,'' Rev. Mod. Phys. 21 (1949) 447--450.

\bibitem{Gott1991}
J. R. Gott, ``Closed timelike curves produced by pairs of moving cosmic strings: Exact solutions,'' Phys. Rev. Lett. 66 (Mar, 1991) 1126--1129. \url{https://link.aps.org/doi/10.1103/PhysRevLett.66.1126}.

\bibitem{Morris1988}
M. S. Morris, K. S. Thorne, and U. Yurtsever, ``Wormholes, time machines, and the weak energy condition,'' Phys. Rev. Lett. 61 (Sep, 1988) 1446--1449. \url{https://link.aps.org/doi/10.1103/PhysRevLett.61.1446}.

\bibitem{Lewis2016}
D. Lewis, The Paradoxes of Time Travel, ch. 26, pp. 357--369. John Wiley \& Sons, Ltd, 2016.

\bibitem{Deutsch1991}
D. Deutsch, ``Quantum mechanics near closed timelike lines,'' Physical Review D 44 no. 10, (1991) 3197--3217.

\bibitem{Pienaar2013}
J. L. Pienaar, T. C. Ralph, and C. R. Myers, ``Open timelike curves violate heisenberg's uncertainty principle,'' Phys. Rev. Lett. 110 (Feb, 2013) 060501. \url{https://link.aps.org/doi/10.1103/PhysRevLett.110.060501}.

\bibitem{AaronsonWatrous2009}
S. Aaronson and J. Watrous, ``Closed timelike curves make quantum and classical computing equivalent,'' Proceedings of the Royal Society A 465 no. 2102, (2009) 631--647, arXiv:0808.2669 [quant-ph].

\bibitem{AaronsonEtAl2024}
S. Aaronson, M. Bavarian, T. Cubitt, S. Grewal, G. Gueltrini, R. O'Donnell, and M. Raat, ``Computability theory of closed timelike curves,'' 2024. \url{https://arxiv.org/abs/1609.05507}.

\bibitem{Brun2003}
T. A. Brun, ``Computers with closed timelike curves can solve hard problems efficiently,'' Foundations of Physics Letters 16 no. 3, (2003) 245--253. \url{http://dx.doi.org/10.1023/A:1025967225931}.

\bibitem{Bennett2009}
C. H. Bennett, D. Leung, G. Smith, and J. A. Smolin, ``Can closed timelike curves or nonlinear quantum mechanics improve quantum state discrimination or help solve hard problems?,'' Phys. Rev. Lett. 103 (Oct, 2009) 170502. \url{https://link.aps.org/doi/10.1103/PhysRevLett.103.170502}.

\bibitem{BrunWilde2011}
T. A. Brun and M. M. Wilde, ``Perfect state distinguishability and computational speedups with postselected closed timelike curves,'' Foundations of Physics 42 no. 3, (2011) 341--361. \url{http://dx.doi.org/10.1007/s10701-011-9601-0}.

\bibitem{BrunWildeWinter2013}
T. A. Brun, M. M. Wilde, and A. Winter, ``Quantum state cloning using deutschian closed timelike curves,'' Phys. Rev. Lett. 111 (Nov, 2013) 190401. \url{https://link.aps.org/doi/10.1103/PhysRevLett.111.190401}.

\bibitem{HaydenPreskill2007}
P. Hayden and J. Preskill, ``Black holes as mirrors: quantum information in random subsystems,'' Journal of High Energy Physics 2007 no. 09, (2007) 120--120. \url{http://dx.doi.org/10.1088/1126-6708/2007/09/120}.

\bibitem{YoshidaKitaev2017}
B. Yoshida and A. Kitaev, ``Efficient decoding for the hayden-preskill protocol,'' 2017. \url{https://arxiv.org/abs/1710.03363}.

\bibitem{Lloyd2011PRL}
S. Lloyd, L. Maccone, R. Garcia-Patron, V. Giovannetti, Y. Shikano, S. Pirandola, L. A. Rozema, A. Darabi, Y. Soudagar, L. K. Shalm, and A. M. Steinberg, ``Closed timelike curves via postselection: Theory and experimental test of consistency,'' Phys. Rev. Lett. 106 (Jan, 2011) 040403. \url{https://link.aps.org/doi/10.1103/PhysRevLett.106.040403}.

\bibitem{Ringbauer2014}
M. Ringbauer, M. A. Broome, C. R. Myers, A. G. White, and T. C. Ralph, ``Experimental simulation of closed timelike curves,'' Nature Communications 5 (2014) 4145.

\bibitem{Lloyd2011PRD}
S. Lloyd, L. Maccone, R. Garcia-Patron, V. Giovannetti, and Y. Shikano, ``Quantum mechanics of time travel through post-selected teleportation,'' Physical Review D 84 no. 2, (2011). \url{http://dx.doi.org/10.1103/PhysRevD.84.025007}.

\bibitem{Huang2026}
Y.-T. Huang, H.-W. Huang, J.-D. Lin, A. Miranowicz, N. Lambert, G.-Y. Chen, F. Nori, and Y.-N. Chen, ``Experimental simulation of postselected closed timelike curves for decoding scrambled quantum information,'' Phys. Rev. Res. 8 (Apr, 2026) 023084. \url{https://link.aps.org/doi/10.1103/tm83-sxpm}.

\bibitem{HorowitzMaldacena2004}
G. T. Horowitz and J. Maldacena, ``The black hole final state,'' Journal of High Energy Physics 2004 no. 02, (2004) 008, arXiv:hep-th/0310281.

\bibitem{GottesmanPreskill2004}
D. Gottesman and J. Preskill, ``Comment on ``the black hole final state'','' Journal of High Energy Physics 2004 no. 03, (2004) 026, arXiv:hep-th/0311269.

\bibitem{Maldacena2016}
J. Maldacena, S. H. Shenker, and D. Stanford, ``A bound on chaos,'' Journal of High Energy Physics 2016 no. 08, (2016) 106, arXiv:1503.01409 [hep-th].

\bibitem{Leone2021}
L. Leone, S. F. E. Oliviero, Y. Zhou, and A. Hamma, ``Quantum Chaos is Quantum,'' Quantum 5 (May, 2021) 453. \url{https://doi.org/10.22331/q-2021-05-04-453}.

\bibitem{Mi2021}
X. Mi, P. Roushan, C. Quintana, S. Mandr\`a, J. Marshall, C. Neill, F. Arute, K. Arya, J. Atalaya, R. Babbush, J. C. Bardin, R. Barends, J. Basso, A. Bengtsson, S. Boixo, A. Bourassa, M. Broughton, B. B. Buckley, D. A. Buell, B. Burkett, N. Bushnell, Z. Chen, B. Chiaro, R. Collins, W. Courtney, S. Demura, A. R. Derk, A. Dunsworth, D. Eppens, C. Erickson, E. Farhi, A. G. Fowler, B. Foxen, C. Gidney, M. Giustina, J. A. Gross, M. P. Harrigan, S. D. Harrington, J. Hilton, A. Ho, S. Hong, T. Huang, W. J. Huggins, L. B. Ioffe, S. V. Isakov, E. Jeffrey, Z. Jiang, C. Jones, D. Kafri, J. Kelly, S. Kim, A. Kitaev, P. V. Klimov, A. N. Korotkov, F. Kostritsa, D. Landhuis, P. Laptev, E. Lucero, O. Martin, J. R. McClean, T. McCourt, M. McEwen, A. Megrant, K. C. Miao, M. Mohseni, S. Montazeri, W. Mruczkiewicz, J. Mutus, O. Naaman, M. Neeley, M. Newman, M. Y. Niu, T. E. O'Brien, A. Opremcak, E. Ostby, B. Pato, A. Petukhov, N. Redd, N. C. Rubin, D. Sank, K. J. Satzinger, V. Shvarts, D. Strain, M. Szalay, M. D. Trevithick, B. Villalonga, T. White, Z. J. Yao, P. Yeh, A. Zalcman, H. Neven, I. Aleiner, K. Kechedzhi, V. Smelyanskiy, and Y. Chen, ``Information scrambling in quantum circuits,'' Science 374 no. 6574, (Dec., 2021) 1479--1483. \url{http://dx.doi.org/10.1126/science.abg5029}.

\bibitem{YoshidaYao2019}
B. Yoshida and N. Y. Yao, ``Disentangling scrambling and decoherence via quantum teleportation,'' Phys. Rev. X 9 (Jan, 2019) 011006. \url{https://link.aps.org/doi/10.1103/PhysRevX.9.011006}.

\bibitem{Zhuang2019}
Q. Zhuang, T. Schuster, B. Yoshida, and N. Y. Yao, ``Scrambling and complexity in phase space,'' Physical Review A 99 no. 6, (2019). \url{http://dx.doi.org/10.1103/PhysRevA.99.062334}.

\bibitem{Provost1980}
J. P. Provost and G. Vallee, ``Riemannian structure on manifolds of quantum states,'' Communications in Mathematical Physics 76 no. 3, (1980) 289--301.

\bibitem{Kolodrubetz2017}
M. Kolodrubetz, D. Sels, P. Mehta, and A. Polkovnikov, ``Geometry and non-adiabatic response in quantum and classical systems,'' Physics Reports 697 (2017) 1--87, arXiv:1602.01062 [cond-mat.stat-mech].

\bibitem{Cheng2013}
R. Cheng, ``Quantum geometric tensor (fubini-study metric) in simple quantum system: A pedagogical introduction,'' 2013. \url{https://arxiv.org/abs/1012.1337}.

\bibitem{Kang2024}
M. Kang, S. Kim, Y. Qian, P. M. Neves, L. Ye, J. Jung, D. Puntel, F. Mazzola, S. Fang, C. Jozwiak, A. Bostwick, E. Rotenberg, J. Fuji, I. Vobornik, J.-H. Park, J. G. Checkelsky, B.-J. Yang, and R. Comin, ``Measurements of the quantum geometric tensor in solids,'' Nature Physics 21 no. 1, (Nov., 2024) 110--117. \url{http://dx.doi.org/10.1038/s41567-024-02678-8}.

\bibitem{Austrich2022}
J. A. Austrich-Olivares and J. D. Vergara, ``The quantum geometric tensor in a parameter-dependent curved space,'' Entropy 24 no. 9, (2022) 1236. \url{http://dx.doi.org/10.3390/e24091236}.

\bibitem{Bishop2025}
L. G. Bishop, F. Costa, and T. C. Ralph,
``Quantum state tomography on closed timelike curves using weak
measurements,''
Class. Quantum Grav. \textbf{42}, 045018 (2025).
\url{https://doi.org/10.1088/1361-6382/ada90b}

\bibitem{Grover1996}
L. K. Grover, ``A fast quantum mechanical algorithm for database search,'' Proceedings of the 28th Annual ACM Symposium on Theory of Computing (STOC) (1996) 212--219, arXiv:quant-ph/9605043.

\bibitem{Brassard2002}
G. Brassard, P. H\o yer, M. Mosca, and A. Tapp, ``Quantum amplitude amplification and estimation,'' in Quantum Computation and Information, vol. 305 of Contemporary Mathematics, pp. 53--74. American Mathematical Society, 2002. arXiv:quant-ph/0005055.

\bibitem{Aaronson2005}
S. Aaronson, ``Quantum computing, postselection, and probabilistic polynomial-time,'' Proceedings of the Royal Society A 461 no. 2063, (2005) 3473--3482, arXiv:quant-ph/0412187.

\bibitem{Huang2020}
H.-Y. Huang, R. Kueng, and J. Preskill, ``Predicting many properties of a quantum system from very few measurements,'' Nature Physics 16 no. 10, (2020) 1050--1057. \url{http://dx.doi.org/10.1038/s41567-020-0932-7}.

\bibitem{Wisniacki2012}
A. Wisniacki, ``Loschmidt echo,'' Scholarpedia 7 no. 8, (2012) 11687. \url{http://dx.doi.org/10.4249/scholarpedia.11687}.

\bibitem{Morapakula2026}
S. N. Morapakula, ``Closed-Time-Like-Curves,'' Apr., 2026. \url{https://github.com/Qubit1718/Closed-Time-Like-Curve}.

\bibitem{Landsman2019}
K. A. Landsman, C. Figgatt, T. Schuster, N. M. Linke, B. Yoshida, N. Y. Yao, and C. Monroe, ``Verified quantum information scrambling,'' Nature 567 no. 7746, (Mar., 2019) 61--65. \url{http://dx.doi.org/10.1038/s41586-019-0952-6}.







\end{thebibliography}
\end{document}